\documentclass[twocolumn,letterpaper,aps,prc,longbibliography,superscriptaddress,showpacs,floatfix,nofootinbib]{revtex4-1}

\usepackage{graphicx}	
\usepackage{xspace}     
\usepackage{amsmath}
\usepackage[usenames,dvipsnames]{color}

\newcommand{\Npart}{\mbox{$N_{\rm part}$}\xspace}
\newcommand{\Ncoll}{\mbox{$N_{\rm coll}$}\xspace}
\newcommand{\Nch}{\mbox{$N_{\rm ch}$}\xspace}
\newcommand{\Et}{\mbox{${\rm E}_T$}\xspace}

\newcommand{\sqs}{\mbox{$\sqrt{s}$}\xspace}
\newcommand{\sqsn}{\mbox{$\sqrt{s_{_{NN}}}$}\xspace}

\newcommand{\Nqp}{\mbox{$N_{\rm qp}$}\xspace}

\newcommand{\dEt}{\mbox{$dE_T/d\eta$}\xspace}
\newcommand{\dNch}{\mbox{$dN_{\rm ch}/d\eta$}\xspace}

\newcommand{\dNchNormQ}{\mbox{$dN_{\rm ch}/d\eta / (0.5 \Nqp)$}\xspace}

\def\lsim{\raise0.3ex\hbox{$<$\kern-0.75em\raise-1.1ex\hbox{$\sim$}}}
\def\gsim{\raise0.3ex\hbox{$>$\kern-0.75em\raise-1.1ex\hbox{$\sim$}}}
\def\mean#1{\left<#1\right>}

\begin{document}
\title{Tests of constituent-quark generation methods which maintain both the nucleon center of mass and the desired radial distribution in Monte Carlo Glauber models.}

\author{J.~T.~Mitchell, D.~V.~Perepelitsa, M.~J.~Tannenbaum\\ {\it Brookhaven National Laboratory, Upton, NY 11973 USA} \\[0.5pc] P.W.~Stankus\\ {\it Oak Ridge National Laboratory\\ Oak Ridge TN 37831 USA }}

\date{\today}

\begin{abstract}

Several methods of generating three constituent-quarks in a nucleon are evaluated which explicitly maintain the nucleon's center of mass and desired radial distribution and can be used within Monte Carlo Glauber frameworks. The geometric models provided by each method are used to generate distributions over the number of constituent-quark participants (\Nqp) in $p+p$, $d+$Au and Au$+$Au collisions.  The results are compared with each other and to a previous result of \Nqp calculations, without this explicit constraint, used in measurements of \sqsn=200 GeV $p+p$, $d+$Au and Au$+$Au collisions at RHIC.  

\end{abstract}

\pacs{25.75.Dw}  
	
\maketitle


\section{Glauber Models for A$+$B collisions}
In order to calculate the nuclear geometry, i.e. the number of participating nucleons (and more recently constituent-quark participants) in the collision of two nuclei, the Glauber Monte Carlo approach is widely used~\cite{Miller:2007ri}. The nucleons in the two nuclei are taken to be frozen in position in their respective nucleus during the collision where it is also assumed that the nucleons travel along straight line trajectories which are not affected by any nucleon-nucleon (N$+$N) collisions. This enables a high-energy nucleus-nucleus (A$+$B) collision to be simulated in two steps:
\begin{itemize}
\item[{(i)}] Each beam and target nucleus, composed of $A$ and $B$ nucleons respectively is generated by placing the nucleons at random about the position of the nucleus, uniform in $\cos\theta$, $\phi$ and $r^2\times \rho_{N}(r)$ for spherical nuclei, where the nuclear density, $\rho_{N}(r)$, is typically taken as a Fermi (Woods-Saxon) density function~\cite{HofstadterRMP28}:
\begin{equation}
\frac{d^3 {\cal P} } {d^3 r} =\frac{d^3 {\cal P}} {r^2 dr \sin\!\theta d\theta\,  d\phi}=\rho_{N}(r)=\frac{\rho_0}{1 + \exp (\frac{r-c}{a_0})} \label{eq:WS}
\end{equation}
with $c=\{1.18 A^{1/3} - 0.48\}$ fm and the diffusivity $a_0=0.545$ fm for Au. 
\item[{(ii)}] The coordinates of the two nuclei are then shifted at random relative to each other in a plane transverse to the beam axis by a vector $\vec{b}$, the impact parameter, sampling an area much larger than the range of possible impact parameters for the A$+$B collision. A pair of nucleons, one from each nucleus, interact with each other if their projected distance $d$ in the plane transverse to the beam axis satisfies the condition:
\begin{equation}
d\leq \sqrt{\frac{\sigma^{\rm inel}_{N+N}}{\pi}}  \label{eq:nn}
\end{equation} 
where $\sigma^{\rm inel}_{N+N}$ is the inelastic nucleon-nucleon cross section.  
\end{itemize}
The nuclear geometrical parameters such as \Npart, the number of participating nucleons, \Ncoll, the number of binary $N+N$ collisions, etc., can be computed for each trial.   

\subsection{Constituent-quarks}
The approach can be extended one level further down with an additional step:  
\begin{itemize}
\item[{(iii)}] modeling each nucleon by generating the positions of the three constituent-quarks ($q$) within it~\cite{MGMPL8,MuellerNPA527,MorpurgoRNC33}, effectively creating a constituent-quark structure of each nucleus. The A$+$B collision represented as a sum of $q+q$ collisions is then obtained by changing the word ``nucleon'' to ``quark''  and $N+N$ to $q+q$ in step (ii) above.
\end{itemize}
\section{\Et and \Nch distributions and Extreme Independent Models (EIM) }
\label{sec:EIMs}
   Measurements of mid-rapidity transverse energy distributions $d\Et/d\eta$ in $p$+$p$, $d$+Au and Au+Au collisions at \sqsn=200 GeV were recently presented by the PHENIX collaboration~\cite{ppg100}. The transverse energy \Et, a multiparticle variable dominated by soft particles and closely related to the multiplicity of charged particles \Nch~\cite{Adler:2004zn}, is defined as the sum
   \begin{eqnarray}
   \Et&=&\sum_i E_i\,\sin\theta_i\cr
   d\Et(\eta)/d\eta&=&\sin\theta(\eta)\, dE(\eta)/d\eta \quad, 
   \label{eq:ETdef}
   \end{eqnarray}
where $\theta$ is the polar angle, $\eta=-\ln \tan\theta/2$ is the pseudorapidity, $E_i$ is by convention taken as the kinetic energy for baryons, the kinetic energy + 2 $m_N$ for antibaryons, and the total energy for all other particles, and the sum is taken over all particles emitted into a fixed solid angle for each event. 
The PHENIX measurement~\cite{ppg100} showed that the number of constituent-quark participants \Nqp represented the fundamental elements of particle production at mid-rapidity in all 3 systems. 

The data were analyzed in the Extreme Independent Model (EIM) framework in which the effect of the nuclear geometry of the interaction can be calculated independently of the dynamics of particle production, which can be derived from experimental measurements, usually the $p$+$p$ (or $p$+$A$) measurement in the same detector. The nuclear geometry of the collisions is usually calculated in a Glauber Monte Carlo Calculation~\cite{Miller:2007ri} and represented as the relative probability of the number $n$ of fundamental elements of particle production, called weights $w_n$, with $n=1$ or 2 as a minimum, to a maximum value $N_{\rm max}$.  EIMs, which have been popular since the first measurements of particle production in relativistic $p+$A and A$+$A collisions, successfully describe \Nch and \Et distributions in different ranges of c.m. energies \sqsn. A representative sample includes the wounded-nucleon (or nucleon-participant \Npart) model (WNM)~\cite{WNM}, wounded-projectile-nucleon Model (WPNM)~\cite{FtPLB188,E802PLB197,E802PRC63}, additive-quark-model (AQM)~\cite{AQMPRD25} which is equivalent to a wounded-projectile-quark (color-string) model, constituent-quark-participant model (NQP)~\cite{ppg100} and quark-diquark model~\cite{BialasBzdakPLB649} (where the names reflect the fundamental element of particle production, nucleons or constituent-quarks.~\footnote{In the AQM, unlike the other models, a distinction is made between the number of constituent-quark participants and the mechanism of particle production which is by color strings between the quark participants with the restriction of only one color string attached to a quark participant. In an asymmetric A$+$B collision, this restriction limits the number of color-strings to the number of constituent-quarks in the smaller nucleus so is effectively a projectile-quark participant model, while the NQP model allows all the quark participants in both nuclei to emit particles.})

At RHIC (\sqsn$=19.6-200$ GeV), PHOBOS~\cite{Alver-Nch-PRC83} has shown that the WNM works in Au+Au collisions for the total multiplicity, \Nch, over the range $|\eta|<5.4$, while at mid-rapidity, the WNM fails---the multiplicity density per participant pair, $\mean{d\Nch/d\eta}/(\Npart/2)$, increases with increasing number of participants, in agreement with previous PHENIX results~\cite{Adcox:2000sp,Adcox:2001ry,Adler:2004zn}. 
Additionally, it was shown using PHOBOS Au+Au data~\cite{EreminVoloshinPRC67,NouicerEPJC49} and discussed for other data~\cite{DeBhattPRC71} that the mid-rapidity $\mean{d\Nch/d\eta}$ as a function of centrality in Au+Au collisions is linearly proportional to the number of constituent-quark participants (NQP); however for symmetric systems this cannot be distinguished from the number of color-strings, the AQM~\cite{Bialas2008}. 

PHENIX~\cite{ppg100} then demonstrated, using mid-rapidity \Et distributions at \sqsn=200 GeV in the asymmetric $d$+Au system, as well as $p$+$p$ and Au+Au collisions, that the asymmetric $d$+Au measurement, which is crucial in distinguishing the color-string (AQM) from NQP models, clearly rejects the AQM and agrees very well with the NQP model. The NQP model also explained that the two-component ansatz,  \mbox{$d\Et/d\eta\propto (1-x) \Npart/2 + x \Ncoll$}, which has been used to describe \Et and charged-multiplicity (\Nch) distributions as a function of centrality, works because the particular linear combination of \Npart and \Ncoll is an empirical proxy for \Nqp and not because the \Ncoll implies a hard-scattering component in \Et or \Nch distributions (which is known to be absent in $p+p$ collisions~\cite{NA5ETnojets,UA2-hard-soft-PLB165}).  
\subsection{Previous methods of generating the positions of the constituent-quarks}
\label{sec:previous}
The first 3 calculations which showed that \Nch was linearly proportional to \Nqp~ \cite{EreminVoloshinPRC67,NouicerEPJC49,DeBhattPRC71} only studied Au$+$Au collisions and simply generated three times the number of nucleons according to the Au radial disribution, Eq.~\ref{eq:WS}, called them constituent-quarks and let them interact with the conventional constituent $q+q$ cross section $\sigma^{\rm inel}_{q+q}=\sigma^{\rm inel}_{N+N}/9$, e.g $\sigma^{\rm inel}_{q+q}$=41mb/9=4.56 mb at \sqsn=130 GeV~\cite{EreminVoloshinPRC67}. 

\begin{figure}[hbt] 
      \centering
 \includegraphics[width=0.88\linewidth]{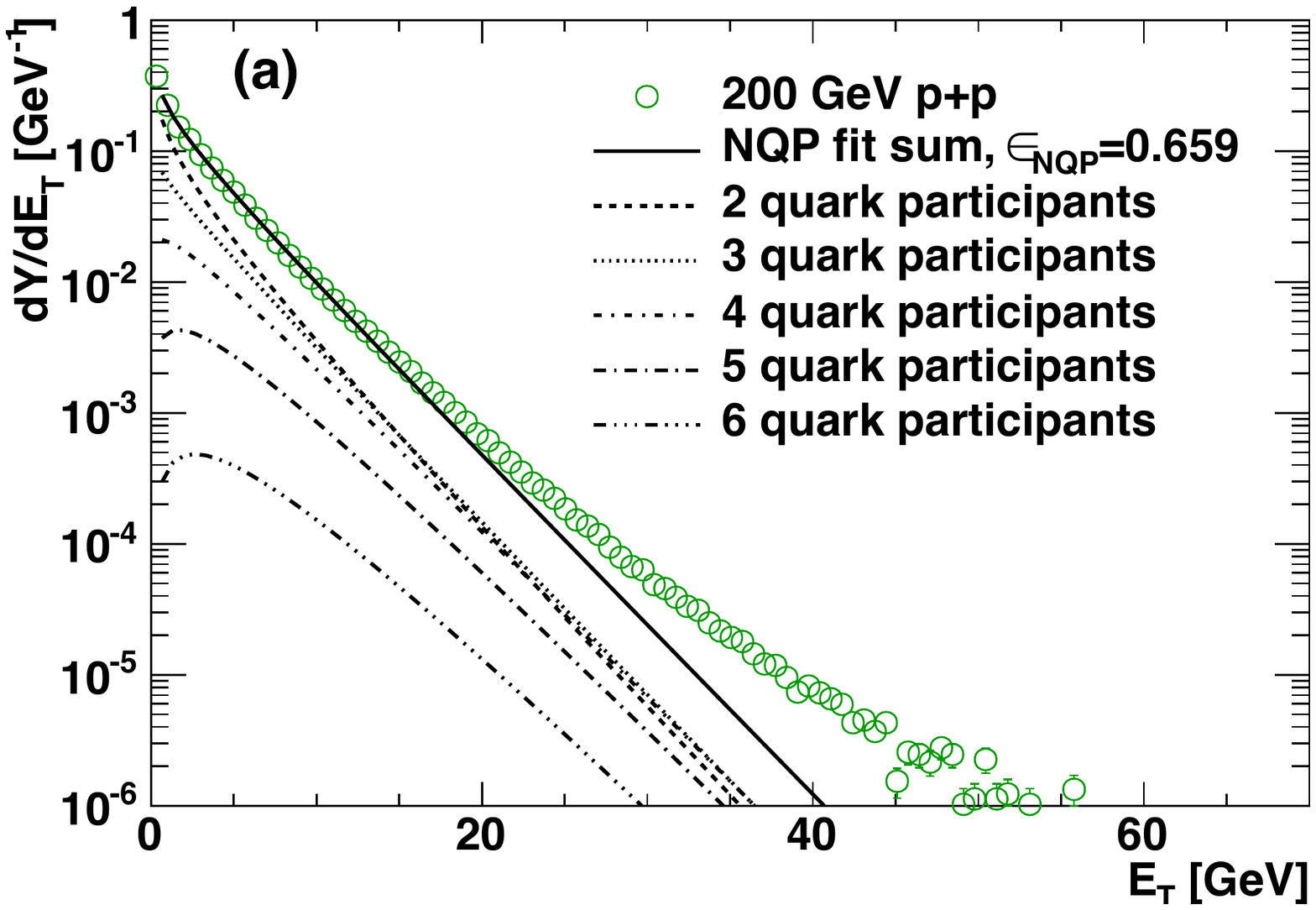}\vspace*{-0.4pc}
      \includegraphics[width=0.91\linewidth]{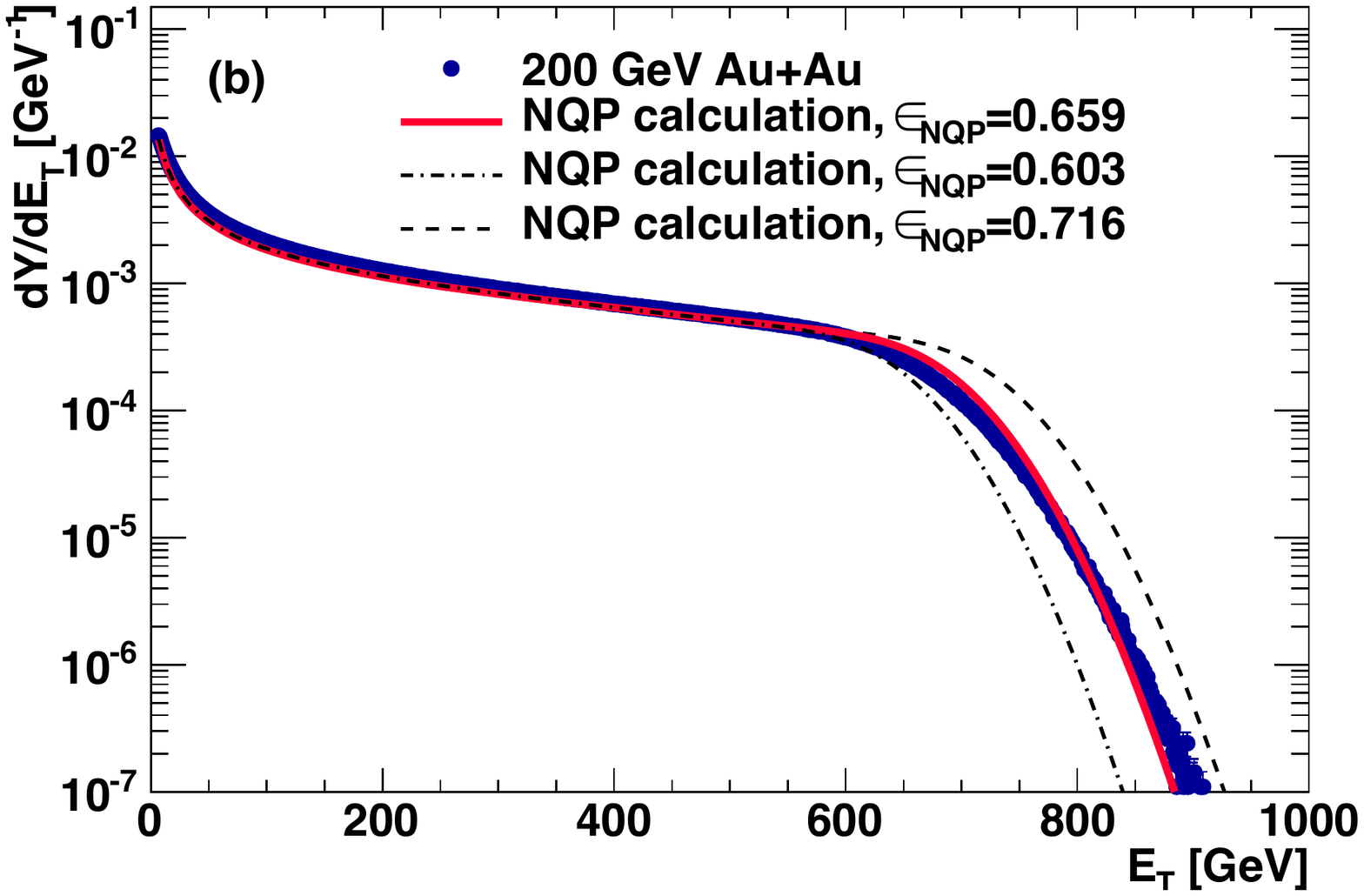}\vspace*{-0.5pc} 
       \includegraphics[width=0.88\linewidth]{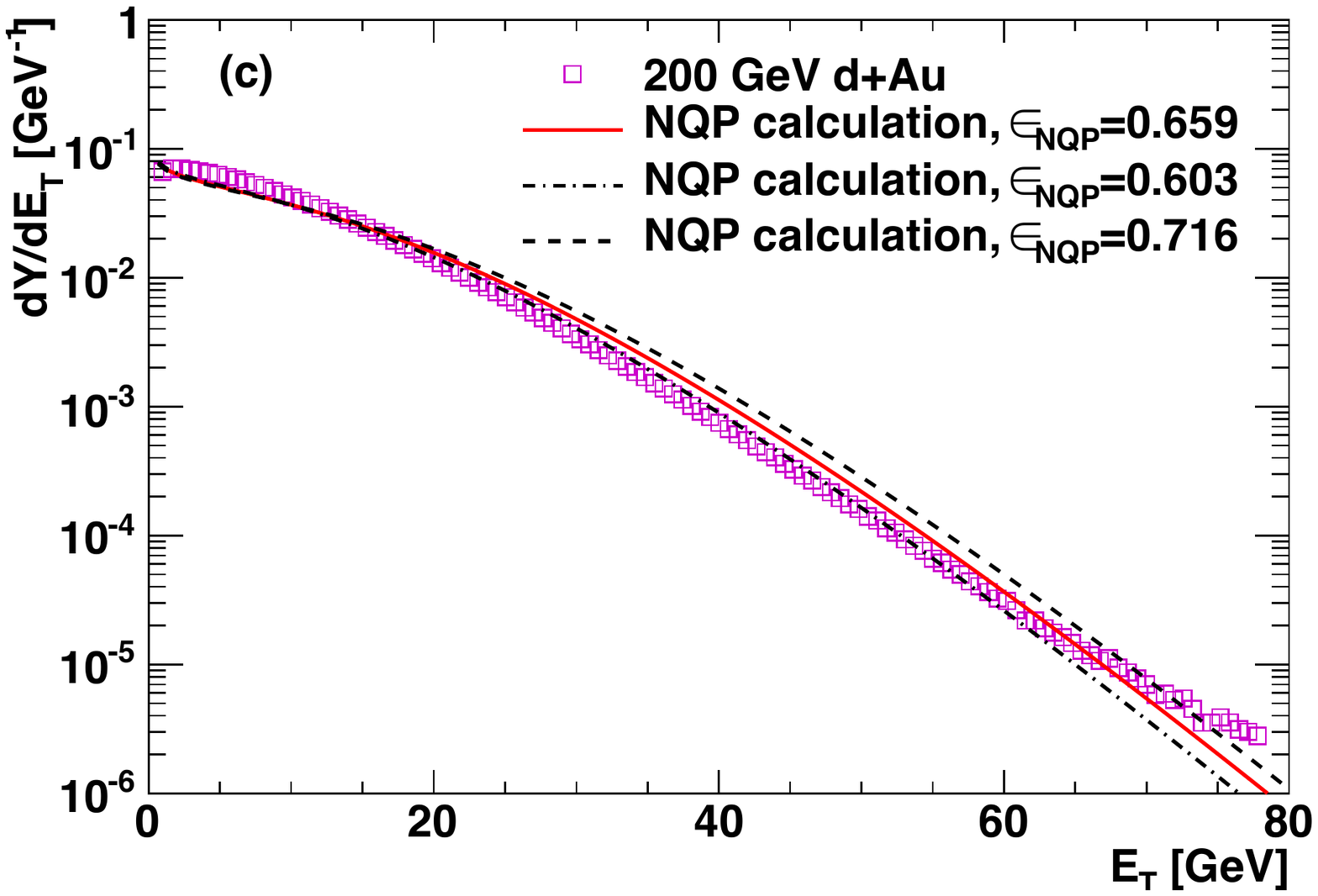}
      \caption[]{PHENIX2014~\cite{ppg100} method for $\Et\equiv d\Et/d\eta|_{y=0}$ distributions at $\sqsn=200$ GeV:
      a) Deconvolution fit to the $p$+$p$ \Et distribution for $\Et<13.3$ GeV with the corrected weights ${w'}^{\rm NQP}_{i}$ with $\epsilon_{\rm NQP}=1-p_{0_{\rm NQP}}=0.659$ calculated in the Number of Quark Participants or \Nqp model. Lines represent the properly weighted individual \Et distributions for the underlying 2,3,4,5,6 constituent-quark participants plus the sum.  b) Au+Au compared to the NQP calculations using the central $1-p_0=0.647$ and $\pm 1\sigma$ variations of $1-p_0=0.582,0.712$ for the probability $p_0$ of getting zero \Et on a $p$+$p$ collision with resulting $\epsilon_{\rm NQP}=0.659,0.603,0.716$, respectively.  c) $d$+Au calculation for the same conditions as in (b).}
      \label{fig:PXppg100}
   \end{figure}

The PHENIX2014 method~\cite{ppg100} was different from these \Nqp calculations in that it used the \Et distribution measured in $p+p$ collisions to derive the \Et distribution of a constituent-quark to use as the basis of the calculations of the $d+$Au and Au$+$Au distributions.  In PHENIX2014~\cite{ppg100}, the spatial positions of the three quarks  were generated around the position of each nucleon in the Glauber Monte Carlo calculations for $p+p$, $d+$Au and Au$+$Au collisions 
using the proton charge distribution corresponding to the Fourier transform of the form factor of the proton~\cite{HofstadterRMP28,HofstadterRMP30}:
\begin{equation}
   \rho^{\rm proton}(r) = \rho^{\rm proton}_{0} \times \exp(-ar),
   \label{eq:Hofstadterdipole}
\end{equation}
where $a = \sqrt{12}/r_{m} = 4.27$ fm$^{-1}$ and 
$r_{m}=0.81$ fm is 
the r.m.s radius of the proton weighted according to charge~\cite{HofstadterRMP28} 
\begin{equation}
r_{m}=\int_0^\infty r^2 \times 4\pi r^2 \rho^{\rm proton}(r) dr \qquad .
\label{eq:rmsintegral}
\end{equation}
The corresponding proton form factor is the Hofstadter dipole fit~\cite{HandMillerWilson} now known as the standard dipole~\cite{BernauerMainzPRC90}:
\begin{equation}
G_E(Q^2)=G_M(Q^2)/\mu=\frac{1}{(1+\frac{Q^2}{0.71 {\rm GeV}^2})^2} \label{HofstadterdipoleFF}
\end{equation}
where $G_E$ and $G_M$ are the electric and magnetic form factors of the proton, $\mu$ is its magnetic moment and $Q^2$ is the four-momentum-transfer-squared of the scattering.
The inelastic $q+q$ cross section $\sigma^{\rm inel}_{q+q}=9.36$mb at \sqsn=200 GeV was derived from the $p+p$ \Nqp Glauber calculation by requiring the  calculated $p+p$ inelastic cross section to reproduce the measured $\sigma^{\rm inel}_{N+N}=42$ mb cross section,  and then used for the $d+$Au and Au$+$Au calculations (Fig.~\ref{fig:PXppg100})~\cite{ppg100}. 

Yet another method was used in the quark-diquark calculation~\cite{BialasBzdakPLB649} where the quark-diquark system for a proton was generated so that it maintained the same center of mass as the original proton. The radius of the quark $\vec{r}_q$ was generated according to a gaussian with the r.m.s. charge radius of the proton and then the diquark is put at half the radius of the quark and vectorially opposite, $\vec{r}_q+2\vec{r}_{qq}=0$. This was not the case for the PHENIX2014 method and is generally not the case for Glauber Monte Carlo calculations in large nuclei. However, it was pointed out~\footnote{Thanks to Adam Bzdak and Peter Steinberg for bringing this to our attention.} that not maintaining the proton c.m. might pose a problem. The generated radii of the three quarks, $\vec{r}_1$, $\vec{r}_2$, and $\vec{r}_3$ follow the correct charge distribution (Eq.~\ref{eq:Hofstadterdipole}) about the original proton center, $\vec{r}=0$; but the origin of the generated quark-triplet proton calculated by the vector sum of the three quark positions, $\vec{r}_p\equiv \vec{r}_1 + \vec{r}_2+\vec{r}_3$, will not likely sum to zero, the position of the original proton. Thus, the radial distributions of all the generated quark-triplets computed about the position $\vec{r}_p$ in each case would be different than the true proton radial distribution and rms radius (Eq.~\ref{eq:Hofstadterdipole}).  This latter problem is shared by the quark-diquark method.

To determine the effects, if any, of the issues raised above about the PHENIX2014~\cite{ppg100} method, 
three new methods of generating constituent-quarks around a nucleon are presented which  will keep the origin of each generated quark-triplet the same as the original nucleon and maintain the desired radial charge distribution. The NQP model calculations are then performed using the PHENIX2014 data~\cite{ppg100}.

\section{Improved methods of generating the positions of constituent-quarks around a nucleon}
Here we describe our operational methods for generating the
positions of constituent-quarks within a nucleon.  Without delving
into the full information on nucleon internal structures as 
revealed in deep inelastic scattering, we construct our model
of quark positions by simply assuming that their spatial distributions
follow the profile of the charge distribution of the proton, relative
to the system's center of mass, as measured through elastic electron 
scattering ~\cite{HofstadterRMP28, HofstadterRMP30}.  Generally, then, to simulate a set of $n$ 
constituents we need a procedure to produce $n$ random vectors
$( \vec{x}_{1},...,\vec{x}_{n} )$ subject to the following conditions. 

\medskip
(1)~Assuming all the constituents to have the same mass, the
sum of their positions $\sum \vec{x}_{i}$ must be identical for
all instantiations of $( \vec{x}_{1},...,\vec{x}_{n} )$, to respect
the fixed, given position of the nucleon center of mass; for
simplicity we can require $\sum \vec{x}_{i}=0$ for any one
nucleon's constituents within its coordinate system;

\smallskip
(2)~The distribution of $\vec{x}_{i}$ for each individual constituent 
should follow some function $\rho(\vec{x})$ which corresponds
to the measured charge distribution of the proton around its
center of mass.

\medskip
\noindent

Elastic scattering measurements on a charge cloud constrain
the (charge-averaged) individual spatial distributions of the
cloud's constituents around the center of mass; but they do
not constrain the correlations among the constituents. 
Accordingly, we will describe and test here a variety of procedures
for generating the $n$ vectors $( \vec{x}_{1},...,\vec{x}_{n} )$ which
satisfy (1)~and (2)~above but which span a wide range in 
correlation behavior.  In all methods described below we take the masses 
and electrical charges of the constituents to be equal, for simplicity;
and we follow these prescriptions for all methods:

\medskip
{\bf Notation:} The magnitude of a position vector is denoted by $r \equiv |\vec{x}|$,
and $r_{i} \equiv |\vec{x}_{i}|$.  The unit vector $\hat{u}$ represents a random direction
chosen uniformly over the sphere.  For each method we assume a general
number of constituents $n$; though we focus on the $n=3$ case for
nucleons in this paper we also investigate the effect of 2,4 and 5 
constituent-quarks in the PHENIX2014~\cite{ppg100} method to reflect, for instance, the presence of sea quarks or gluons. 

\smallskip
{\bf Parent distribution:} For the charge distribution satisfied by the generated quark-triplets about their center of mass, the goal is to reproduce Eq.~\ref{eq:Hofstadterdipole}.

\medskip
We note that none of the previous Glauber Monte Carlo calculations cited in Sec.~\ref{sec:previous} satisfied both conditions (1) and (2) above;  and we believe that this report is the first to do so with comparison to measured $\Et$ distributions in $p+p$, $d+$Au and Au$+$Au collisions.  As we shall see, these
departures from the desired radial distribution (Eq.~\ref{eq:Hofstadterdipole}) are vanishing in the limit of large $n$, but for 
$n=3$, as in constituent-quarks in a nucleon, the differences between
the present and previous constructions are small but informative~\cite{Adare:2015bua}.
\subsection{Planar Polygon}
\label{sec:MJT}
Generate one quark at $(r,0,0)$ with $r$ drawn from $r^2 \rho^{\rm proton}(r)$. Then instead of generating $\cos\theta$ and $\phi$ at random and repeating for the two other quarks as was done by PHENIX2014~\cite{ppg100}, imagine that this quark lies on a ring of radius $r$ from the origin and  place the two other quarks on the ring at angles spaced by $2\pi/3$ radians.  Then randomize the orientation of the 3-quark ring spherically symmetric about the origin.  This guarantees that the radial density distribution is correct about the origin and the center of mass of the three quarks is at the origin but leaves the three-quark-triplet on each trial forming an equilateral triangle on the plane of the ring which passes through the origin.   

An evident problem with this model is that it introduces a correlation because all quarks on a given trial have the same radius from the center of the proton. An advantage of this model is that it can be easily applied to any number, $n$, of quarks distributed around the ring of radius $r$ at angles spaced by $2\pi/n$ radians.
\subsection{Explicit Joint distribution}
\label{sec:PWS}
In this method we construct the joint distribution explicitly to 
satisfy conditions (1)~and (2)~above, and to be symmetric among
the constituents, but otherwise to have minimal correlations
between them.  With the use of an auxiliary probability 
distribution function $f(\vec{x})$ we write the joint distribution
over all $n$ vectors simply as

\begin{equation}
P( \vec{x}_{1},...,\vec{x}_{n} ) = 
f( \vec{x}_{1} ) \, f( \vec{x}_{2} ) \, ... \, f( \vec{x}_{n} ) \, \delta \left( \sum \vec{x}_{i}/n \right)
\label{eq:joint_basic}
\end{equation}

\noindent
where the Dirac delta $\delta()$ insures the center-of-mass condition but
the constituents are otherwise as independent as possible.

To get an operational procedure from this definition we need to specify
two things: (I)~how do we choose the auxiliary distribution $f(\vec{x})$ such that
the singles distribution of each $\vec{x}_{i}$ follows $\rho(\vec{x})$; and
(II)~given an $f(\vec{x})$, how do we choose a set $( \vec{x}_{1},...,\vec{x}_{n} )$
which will follow the joint distribution in Eq.~\ref{eq:joint_basic}.  We answer
these as follows.

\medskip
{\bf (I})~Let $f^{ [k] }(\vec{x})$ be defined as the $k^{\mathrm{th}}$-order 3-D
convolution of $f(\vec{x})$ with itself; e.g.
$f^{ [2] }(\vec{x}) = f(\vec{x}) \, \circ \circ \circ \, f(\vec{x})$.  Then it can
be shown from Eq.~\ref{eq:joint_basic} that for the singles distribution of
any individual $\vec{x}_{i}$ to follow $\rho(\vec{x})$ the auxiliary
function $f()$ must satisfy

\begin{equation}
\rho(\vec{x}) = f(\vec{x}) \, f^{[n-1]}(-\vec{x}) \qquad .
\label{eq:joint_implicit}
\end{equation}

In practice it may not be straightforward to invert Eq.~\ref{eq:joint_implicit}
and determine$f()$ given $\rho()$; instead it may be sufficient to
use an auxiliary $f(\vec{x})$ determined by trial and error to match the
desired $\rho(\vec{x})$.  For this paper the implementation is with such a
trial and error determination; the auxiliary function defined with two 
parameters $b$ and $c$
\begin{equation}
f(\vec{x}) = \exp -(b \, r /r_{0}) \; \left[ 1 + \left( \frac{r}{c \, r_{0}} \right) \right]
\label{eq:joint_auxiliary}
\end{equation}
with $c=3.9$ and $b=0.91$ and $r_0=r_m/\sqrt{12}$ will result in a singles distribution that
matches the Hofstadter profile~\cite{HofstadterRMP30} Eq.~\ref{eq:Hofstadterdipole}
to within a few percent out to $r < 10 r_0=2.3$fm.  
\medskip

{\bf (II)}~Once the auxiliary function $f()$ is chosen, how do we
operationally generate random sets of position vectors
$(\vec{x}_{1},...,\vec{x}_{n})$, distributed according to the
probability density in Eq.~\ref{eq:joint_basic}.  One simple,
all-purpose approach to generating random variables 
according to any given distribution is {\em rejection 
sampling}.  

In its simplest form, one can calculate the center of mass of the 
generated vectors, $x_{CM}=\Sigma \vec{x}_{i}/n$  
and keep the sample if and only if $x_{CM}$ is within some
tolerance limit of zero, to enforce the effect of the delta
function $\delta( \Sigma \vec{x}_{i}/n )$. 
This method is numerically inefficient as only a small fraction 
of generated sets will be kept; and becomes less efficient the stricter 
the tolerance for $x_{CM}\rightarrow 0$.

\medskip
Fortunately, we can use rejection sampling in a much more
numerically efficient algorithm, which will keep on the
order of 1/5--1/10 of the initially generated selections
and without degrading with increasing $n$, vis:

\smallskip
\noindent
(1)~generate all but one of the vectors $\vec{x}_{i}$
independently, each according to $f(\vec{x})$; which one
of the list is not chosen here is unimportant, let's 
suppose it is $\vec{x}_1$; then

\smallskip
\noindent
(2)~calculate the value of the remaining vector as
the negative sum of all the previously chosen ones,
{\em e.g.} $\vec{x}_{1} = - ( \vec{x}_{2} + ... + \vec{x}_{n} )$
to enforce the center of mass at zero; then

\smallskip
\noindent
(3)~keep this sample if and only if a new random number
chosen uniformly on $[0,1]$ is less than or equal to the probability
density of the final vector, $f(\vec{x}_{1})$; otherwise reject the sample and try again.
\subsection{Recentered approach with empirical radial distribution}
\label{sec:DVPmodel}
\begin{figure}[!h] 
      \centering
      \includegraphics[width=1.00\linewidth]{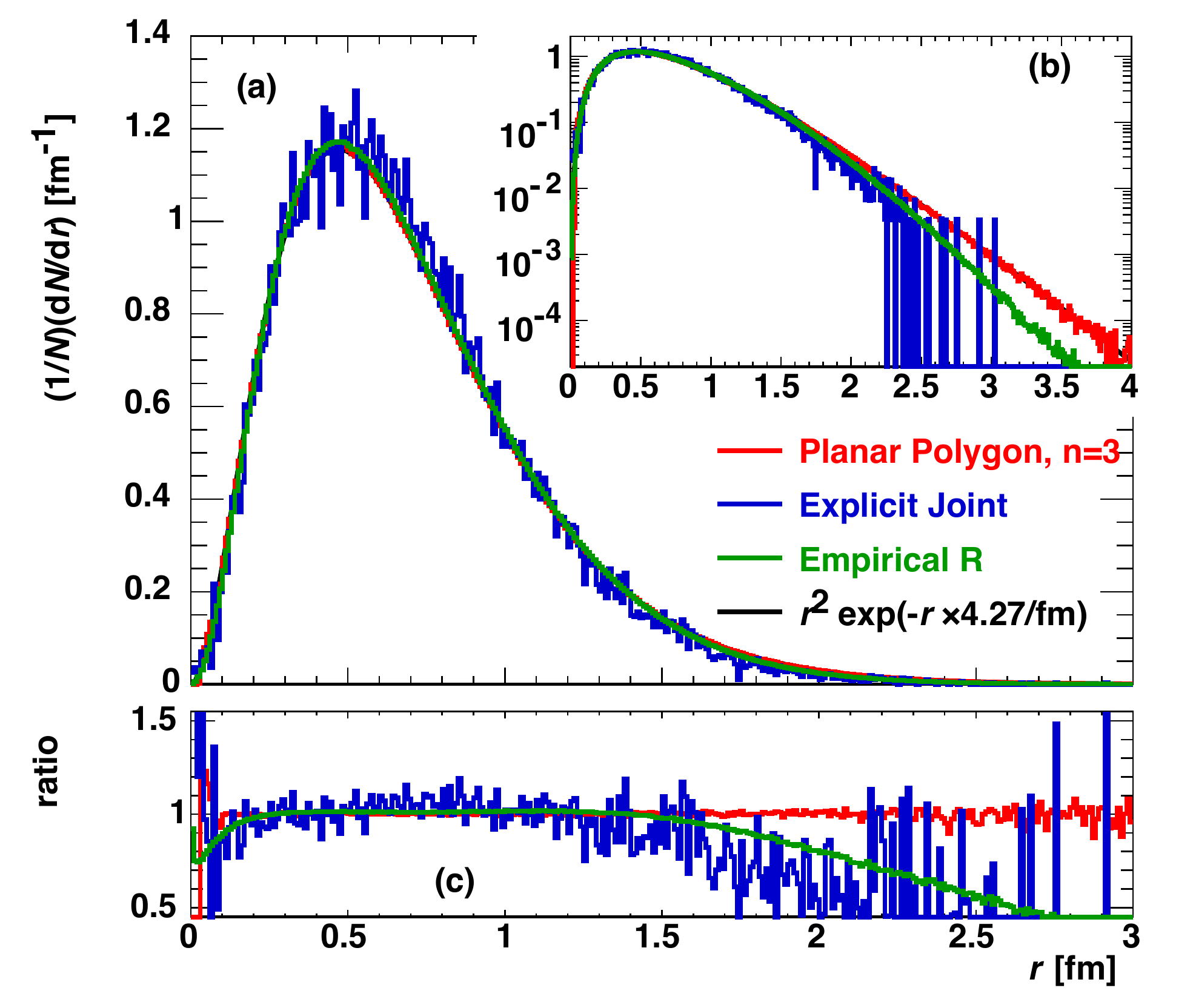} 
      \caption{a) Radial distribution $d{\cal P}/dr$ about the c.m. of the generated quark-triplets as a function of $r$ [fm]  for the three methods:  \mbox{({\color{red}red})} Planar Polygon (Sec.~\ref{sec:MJT}); \mbox{({\color{blue}blue})} Explicit Joint (Sec.~\ref{sec:PWS}); \mbox{({\color{green}green})} Empirical Recentered (Sec.~\ref{sec:DVPmodel}); compared to  $r^2\rho^{\rm proton}(r)$ from Eq.~\ref{eq:Hofstadterdipole} (black), with a semi-log plot shown as the inset (b). (c)  Ratio of the generated distributions to Eq.~\ref{eq:Hofstadterdipole}. The Planar Polygon distribution is identical to $r^2\rho^{\rm proton}(r)$ so it obscures the black line. }
      \label{fig:3radialcm}
   \end{figure}

In this method, the three constituent-quark positions are drawn independently from an auxiliary function $f(r)$, 
and then the center of mass of the generated three-quark system is re-centered to the original nucleon position. The empirical function, $f(r)$ (Eq.~\ref{eq:empirical}), is chosen such that the resulting radial quark distribution $\rho(r)$ with respect to the center of mass (i.e. after re-centering) reproduces the proton charge distribution, $\rho^{\rm proton}(r)$  (Eq.~\ref{eq:Hofstadterdipole}), the Fourier transform of the proton form factor, as measured in electron-proton elastic scattering~\cite{HofstadterRMP30} (Fig.~\ref{fig:3radialcm}). For the results presented here, this function  was chosen to be\footnote{This function was derived by D.~V.~Perepelitsa and used in Ref.~\cite{Adare:2015bua}.} 
\begin{equation}
\begin{split}
   f(r)= r^2 \rho^{\rm proton}(r) (1.21466-1.888r+2.03r^{2}) \\
   \times (1+1.0/r - 0.03/r^{2}) (1+0.15r) 
\end{split}
\label{eq:empirical}
\end{equation}
where $r$ is the radial position of the quark in fm. The 
polar and azimuthal positions of each quark are generated uniformly in $\cos\theta$ and $\phi$ to achieve a 
spherically symmetric distribution.  Once all of the quark coordinates are 
determined, the three quark system is shifted so that the center of mass matches the center position of the nucleon.  

This function was derived through an iterative, empirical approach. For a given test function $f^{\rm test}(r)$, the resulting radial distribution $\rho^{\rm test}(r)$ was compared to the desired distribution $\rho^{\rm proton}(r)$ in Eq.~ \ref{eq:Hofstadterdipole}. The ratio of $\rho^{\rm test}(r) / \rho^{\rm proton}(r)$ was parameterized with a polynomial function of $r$ or $1/r$, and the test function was updated by multiplying it with this parametrization of the ratio. Then, the procedure was repeated 
until the ratio $\rho^{\rm test}(r) / \rho^{\rm proton}(r)$ was sufficiently close to unity over a wide range of $r$ values. The resulting functional form for $f(r)$ is given above in Eq.~\ref{eq:empirical}. A future determination of $f(r)$ may yield an incrementally better agreement between the resulting $\rho(r)$ and the desired Hofstadter standard-dipole form, $\rho^{\rm proton}(r)$ (Eq.~\ref{eq:Hofstadterdipole}), but we believe the present form is close enough for our practical purposes. 

\begin{figure}[!h]
 \centering
 \includegraphics[width=0.99\linewidth]{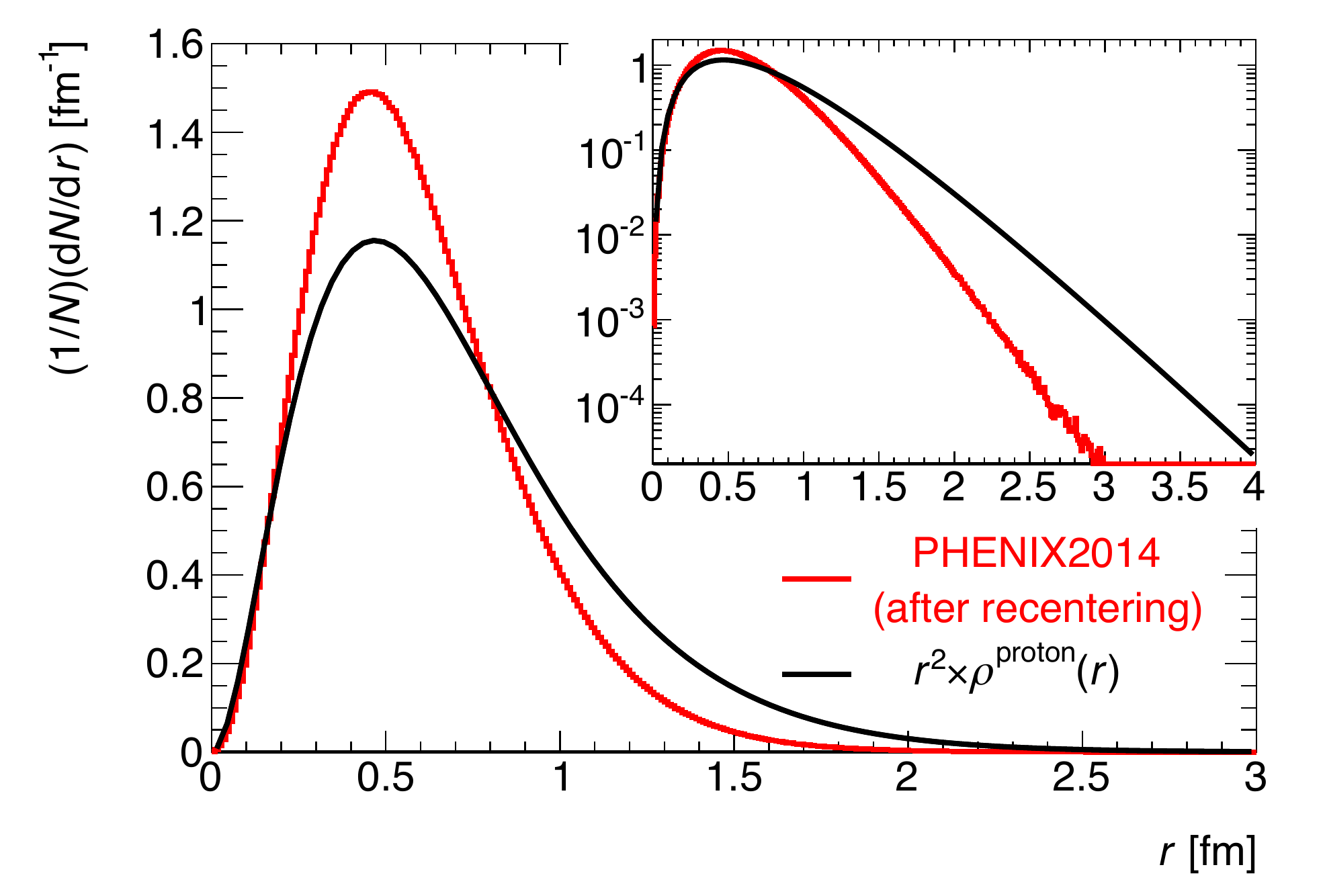}\vspace*{-1pc}
 \caption{Radial distribution $d{\cal P}/dr$ about the c.m. of the generated quark-triplets  as a function of $r$ [fm] for the PHENIX2014~\cite{ppg100} method compared to $r^2\rho^{\rm proton}(r)$ from Eq.~\ref{eq:Hofstadterdipole}.}
 \label{fig:ppg100radial}\vspace*{-0.5pc} 
\end{figure}

This method is conceptually the most similar to the previous PHENIX2014~\cite{ppg100} method of generating constituent-quark systems, which was described in Section~\ref{sec:previous}. That method effectively followed the approach defined here but used an auxiliary function equal to the desired proton charge distribution, $f(r) = \rho^{\rm proton}(r)$ (Eq.~\ref{eq:Hofstadterdipole}). Thus, the resulting $\rho(r)$ with respect to the center of mass (i.e. after re-centering) in the PHENIX2014~\cite{ppg100} method was different from $\rho^{\rm proton}(r)$ (Fig. \ref{fig:ppg100radial}). 

\subsection{Comparison of the three new methods}
 Fig.~\ref{fig:3radialcm} shows the resulting radial distributions in each of the three new methods compared to $r^2\rho^{\rm proton}(r)$ from Eq.~\ref{eq:Hofstadterdipole}.  The Planar Polygon (Sec.~\ref{sec:MJT}) radial distribution $d{\cal P}/dr$ about the c.m. of the generated quark-triplet is  identical to the input proton charge distribution Eq.~\ref{eq:Hofstadterdipole} by construction. However, it has the problem that all 3 quarks in each proton have the same radius from the center and lie in a plane, which passes through the c.m., on the points of an equilateral triangle, an unlikely correlation in nature. The Explicit Joint distribution method (Sec.~\ref{sec:PWS}) follows Eq.~\ref{eq:Hofstadterdipole} to $r\approx 1.5$ fm, almost as well as noted in Sec.~\ref{sec:PWS}, but has issues at small $r$ and is relatively inefficient to generate as shown by the large statistical fluctuations. The Empirical Recentered approach (Sec.~\ref{sec:DVPmodel}) follows Eq.~\ref{eq:Hofstadterdipole} to $r\approx2$ fm, is straightforward to generate and is spherically symmetric, so is likely the best of the three new methods. 
We further note that, unlike the Explicit Joint distribution method (Sec.~\ref{sec:PWS}), each application of this method results in a usable three-quark configuration, making it the more computationally efficient of the two. 

We are now in a position to apply these three methods to Monte Carlo Glauber Model calculations of $p+p$, $d+$Au and Au$+$Au measurements with constituent-quarks as the fundamental elements of particle production,  the NQP  model.\vspace*{-0.5pc}

\subsection{Sensitivity to the number of constituent-quarks per proton}
Although the massive constituent-quarks which form mesons and baryons  (e.g. a proton=$uud$) are relevant for the static properties of hadrons\cite{MGMPL8,MuellerNPA527,MorpurgoRNC33} and soft hadron physics such as \dNch and \dEt,     predominantly composed of particles with $p_T\lsim 1.4$ GeV/c where the $p_T$ distributions of $p$, $K$ and $\pi$ are exponential~\cite{ppg101}, there is often a confusion with massless partons (gluons and current quarks) which are typically only evident in hard scattering with $p_T\gsim 3$ GeV/c where the $p_T$ distributions follow a power-law. Thus, the question typically arises as to whether the NQP model works for an arbitrary number of quarks. 
We have investigated this for the \dNch measurements of Ref.~\cite{Adare:2015bua} using the PHENIX2014 method~\cite{ppg100} of constituent-quark generation which can easily be applied to 2,3,4,5 constituent-quarks, and compared the results to the present methods for 3 consituent-quarks.  
\begin{figure}[!th] 
      \centering
      \includegraphics[width=0.98\linewidth]{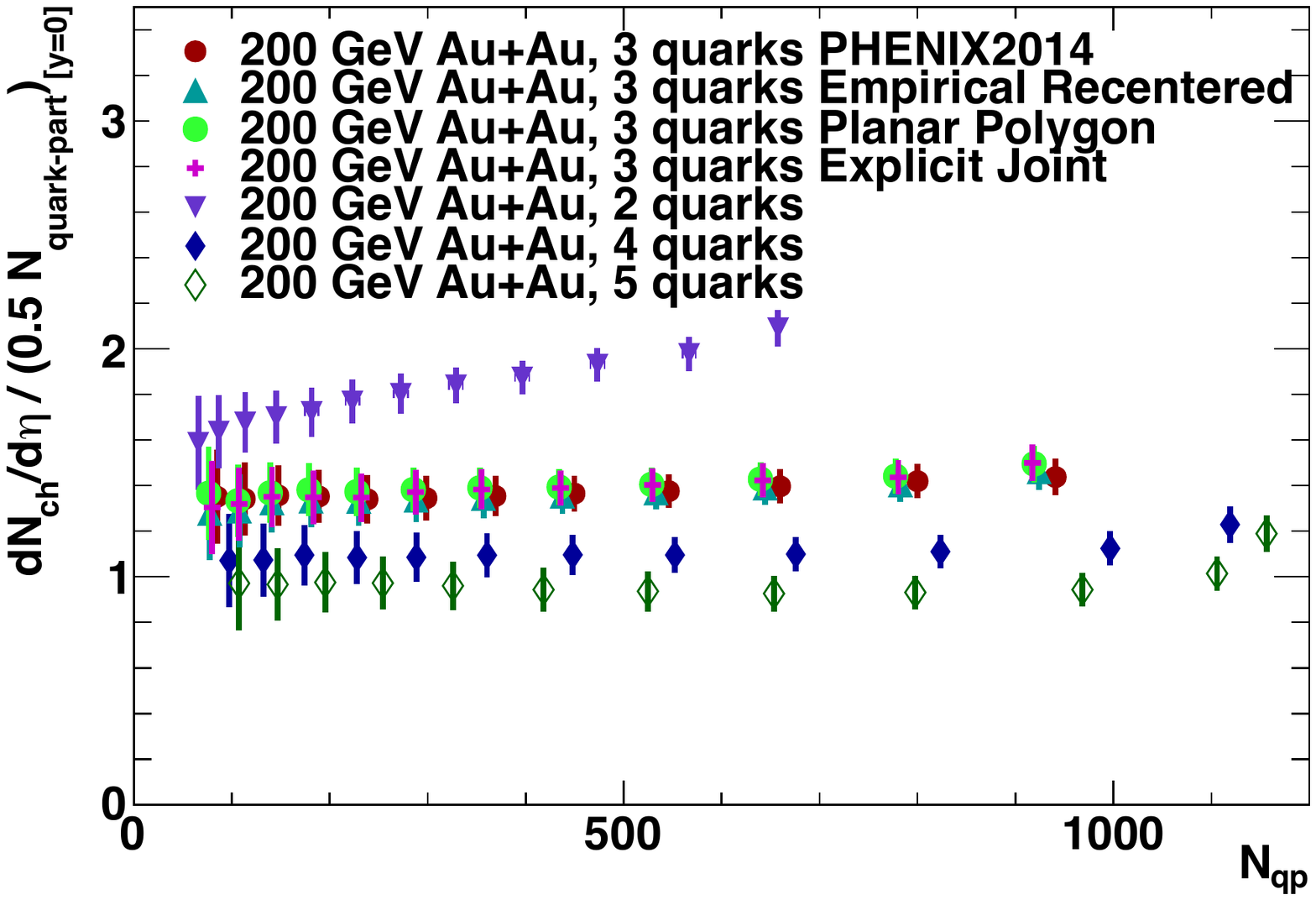}\vspace*{-1pc} 
      \caption{\dNchNormQ at mid-rapidity as a function of \Nqp for \dNch measurements~\cite{Adare:2015bua} in Au$+$Au collisions at \sqsn=200 GeV. The various methods and number of constituent-quarks assumed per proton are given in the legend.  }
      \label{fig:manyquarks}
   \end{figure}
   The results for \dNchNormQ in Au$+$Au at \sqsn=200 GeV, shown in Fig.~\ref{fig:manyquarks}, indicate that 2 constituent-quarks are rejected while 3 constituent-quarks give essentially the same result for all 4 methods, and 4 or 5 constituent-quarks seem to work as well as 3 in this calculation. 

It is also noteworthy that the difference between \dNchNormQ in PHENIX2014~\cite{ppg100} and the present Empirical Recentered approach (Sec.~\ref{sec:DVPmodel}) is $\lsim 2$\%.

\section{Comparison of the new methods to PHENIX2014 results}
The results for NQP calculations for measurements of \Et distributions in $p+p$, $d+$Au and Au$+$Au collisions at \sqsn=200 GeV with the PHENIX2014 constituent-quark generation method~\cite{ppg100} have been shown in Fig.~\ref{fig:PXppg100}. The conclusions were that starting with the \Et distribution of a constituent-quark derived by deconvoluting the $p+p$ \Et distribution according to the number of constituent-quark participants in a $p+p$ collision, both the shape and magnitude of the \Et distributions in $d+$Au and Au$+$Au calculated in the NQP model are in excellent agreement with the measurements. The NQP calculations will now be repeated here using the new methods that keep the origin of each generated quark-triplet the same as the original proton and preserve the generated proton charge distribution.  However, because the calculations are quite detailed, they will be fully described in an  appendix, following the presentation of the results and conclusions. \vspace*{-1pc}

\subsection{NQP Calculations of the $p+p$ distributions}
Calculations of the $q+q$ cross sections $\sigma_{q+q}$ and relative probabilities $w_n$ for $n=2,3,4,5,6$ constituent-quark participants in $p+p$ collisions, for the PHENIX2014 and the three new methods are given in Table~\ref{table:qqwts}. These are derived from Glauber Monte Carlo calculations in each method which vary $\sigma_{q+q}$ until the calculated value for the inelastic cross section in $p+p$ collisions is 42 mb (for \sqsn=200 GeV)~\cite{ppg100}. This is purely nuclear geometry. 

A correction to the weights is then applied to account for the fact that only a fraction $\epsilon_{p+p}$ of $p+p$ collisions produces non-zero \Et in the detector. Gamma distribution parameters $p$ and $b$ of 1 constituent-quark participant are then derived in each method from deconvolution fits to the measured $p+p$ \Et distribution and applied to calculations of \Et distributions in the various collision systems.  

\begin{figure}[hbt] 
      \centering
       \includegraphics[width=0.86\linewidth]{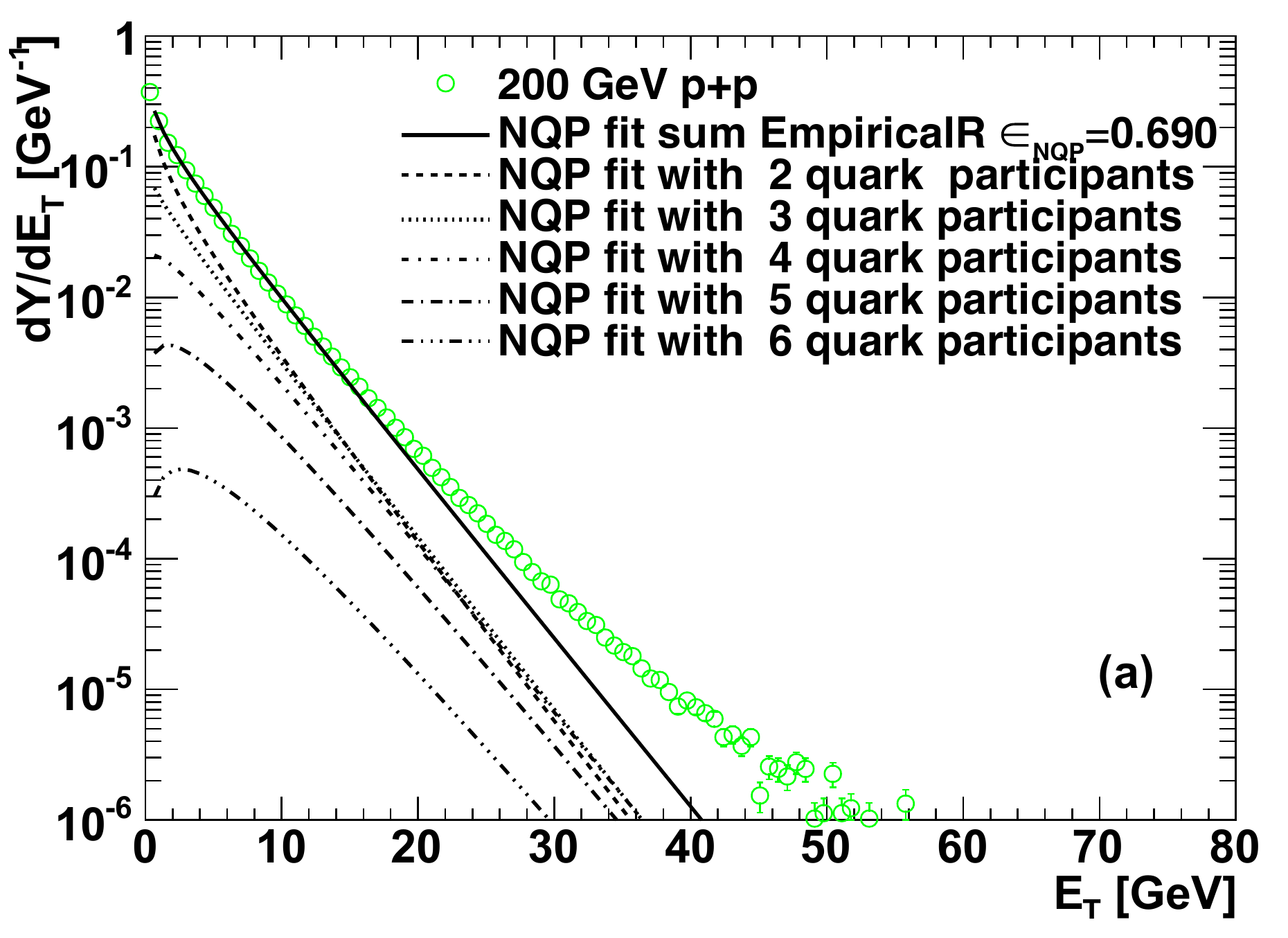}\vspace*{-0.3pc}
     \includegraphics[width=0.86\linewidth]{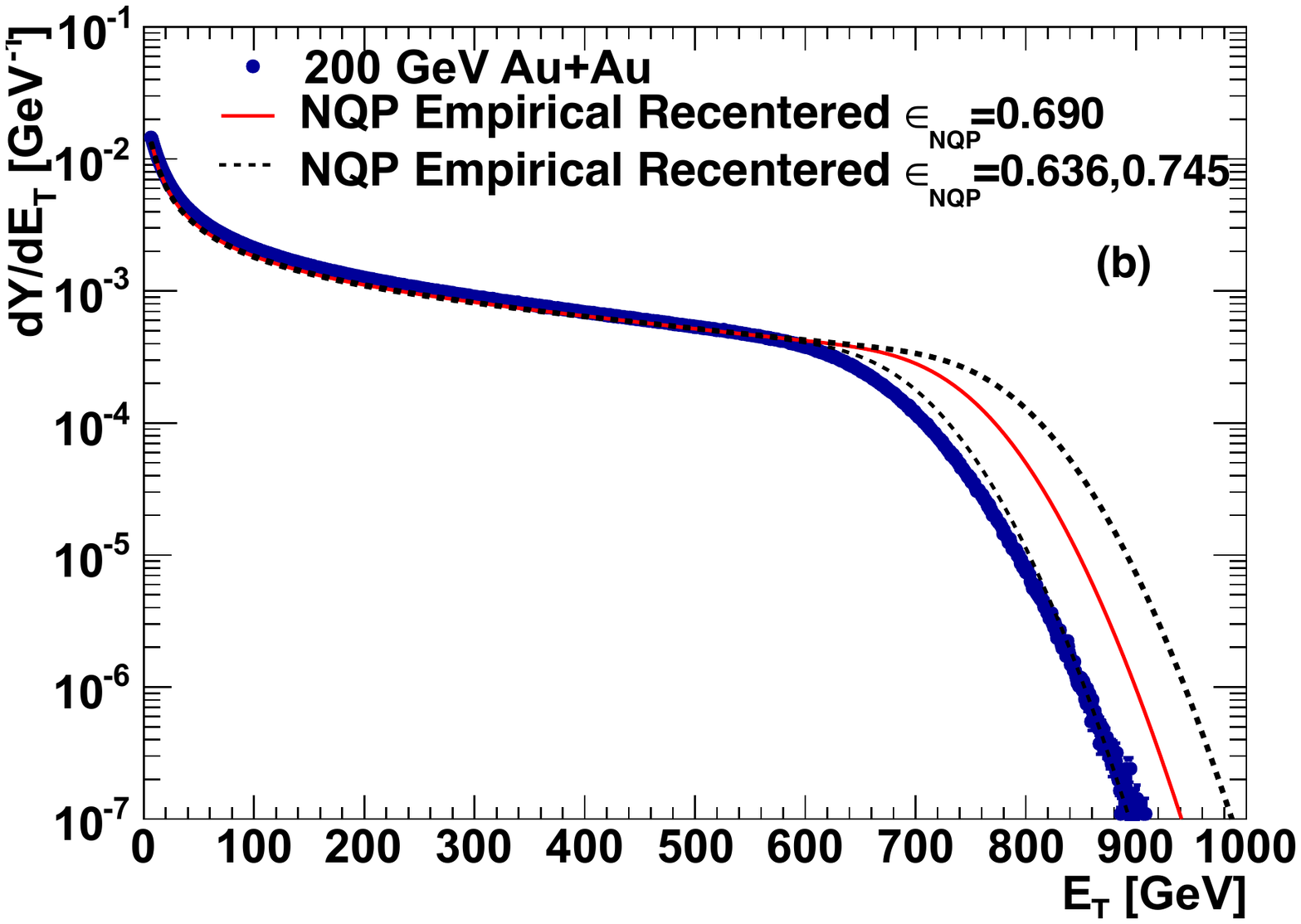}\vspace*{-0.3pc} 
      \includegraphics[width=0.86\linewidth]{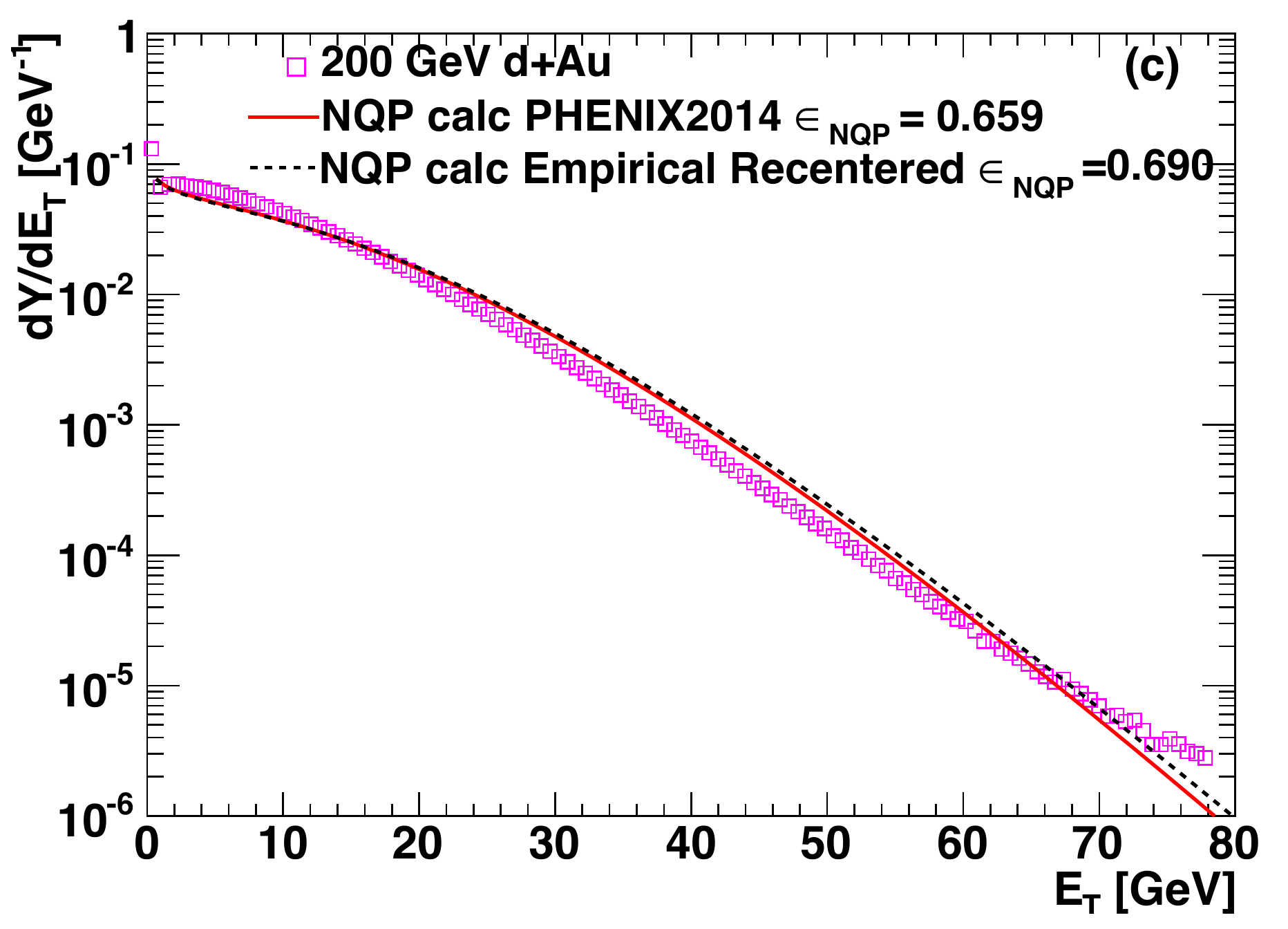} \vspace*{-0.5pc}
      \caption[]{$\Et\equiv d\Et/d\eta|_{y=0}$ distributions at $\sqsn=200$ GeV calculated in the Number of Quark Participants (NQP) model from the Empirical Recentered (Sec.~\ref{sec:DVPmodel}) method with $\epsilon_{\rm NQP}=1-p_{0_{\rm NQP}}=0.690$. a) NQP calculation of the $p$+$p$ \Et distribution for $\Et<13.3$ GeV with the parameters $p$ and $b$ for the \Et distribution of 1 QP from the deconvolution fit to the same data (Table~\ref{table:Gammafits}) where the thin lines shown are the \Et distributions for 2,3,4,5 and 6 QP weighted by $w'_n$ from Table~\ref{table:corrwts} and the thick line is the sum.  b) Au$+$Au compared to the NQP calculations using the central $1-p_0=0.647$ and $\pm 1\sigma$ variations of $1-p_0=0.582,0.712$ for the probability $p_0$ of getting zero \Et on a $p$+$p$ collision~\cite{ppg100} with resulting $\epsilon_{\rm NQP}=0.690,0.636,0.745$, respectively.  c) $d$+Au calculation for the same conditions as in (b) and for PHENIX2014~\cite{ppg100}.} 
      \label{fig:DVP}\vspace*{-2pc}
   \end{figure}

The calculations of the $p+p$ \Et distribution are shown in Fig.~\ref{fig:DVP}a for the Empirical Recentered method and in Figs.~\ref{fig:Poly}a, \ref{fig:m3}a for the Planar Polygon and Explicit Joint  methods, respectively, using the Gamma distribution parameters $p$ and $b$ for the \Et distribution of 1 constituent-quark participant derived from the deconvolution fits to the same data (Table~\ref{table:Gammafits}). The deconvolution fits and and the NQP calculations  of the $p+p$ \Et distribution using the derived $b$ and $p$ parameters are separate calculations, so the excellent agreement of the $p+p$ data with the calculations shows that the three new methods work as well as the PHENIX2014 method~\cite{ppg100} (Fig.~\ref{fig:PXppg100}a). 

\begin{figure*}[!h]
\centering
\includegraphics[width=0.98\linewidth]{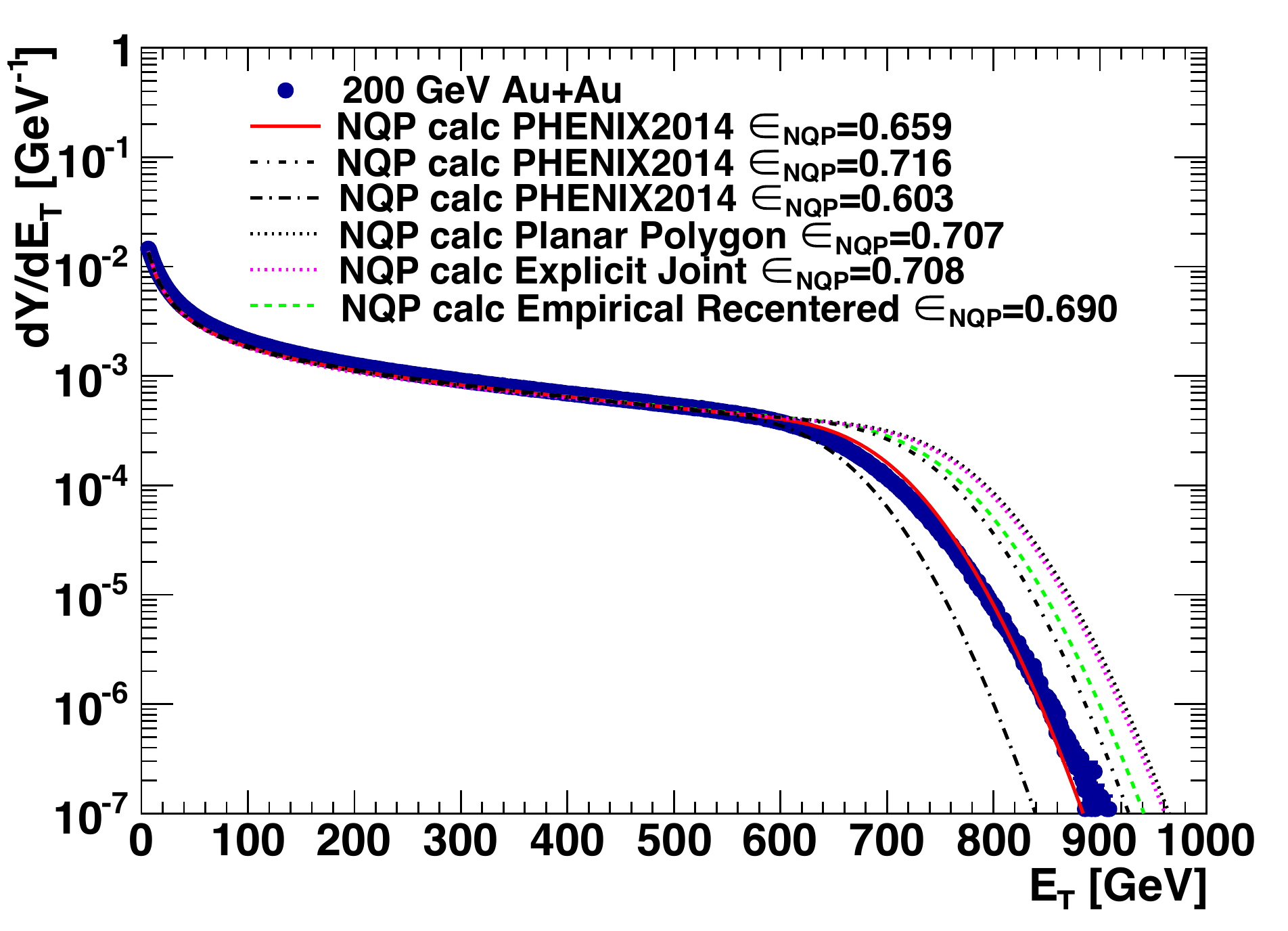} 
\caption{PHENIX2014~\cite{ppg100} \Et distribution in Au$+$Au at $\sqrt{s_{\rm NN}}=200$ GeV compared to the Number of Quark Participants (NQP) model calculations indicated, where $\epsilon_{\rm NQP}$ represents the probability for a constituent-quark participant to give non-zero \Et in the detector. The PHENIX2014 NQP calculation is shown for the central value, $\epsilon_{\rm NQP}=0.659$, and the  $\pm1\sigma$ systematic variations in $\epsilon_{\rm NQP}=0.716,0.603$~\cite{ppg100} as detailed in Fig.~\ref{fig:PXppg100}.  } 
\label{fig:allcalcs}
\end{figure*}
\subsection{NQP Calculations of the Au$+$Au distributions}
    The Au$+$Au calculations are shown in Fig.~\ref{fig:allcalcs} with details in Fig.~\ref{fig:DVP}b and Figs.~\ref{fig:Poly}b, \ref{fig:m3}b which may be compared to Fig.~\ref{fig:PXppg100}b. The Planar Polygon and Empirical Recentered methods agree quite well with the PHENIX2014~\cite{ppg100} calculation and the data but they show 10\% and 7\% more \Et respectively. The surprising result is that the most sophisticated but computing intensive Explicit Joint method comes out virtually on top of the simplest Planar Polygon method rather than in agreement with the Empirical Recentered method. 

More important to note is that the agreement of the new centered methods with the Au$+$Au data is within the $1\sigma$ uncertainty of the Empirical Recentered calculation in Figure~\ref{fig:DVP}b (i.e. the curve of $-1\sigma$ with $\epsilon_{\rm NQP}=0.636$ lies essentially on top of the data) and within $1.5\sigma$ of the Planar Polygon calculation in Figure~\ref{fig:Poly}b.  Also, the new calculations agree with PHENIX 2014 calculation~\cite{ppg100} (Fig.~\ref{fig:allcalcs}) to within $1.2\sigma$ of its uncertainty for the Empirical Recentered method and $1.8\sigma$ for the Planar Polygon method.\vspace*{-1.0pc}  

\subsection{d+Au and AQM vs NQP}
One of the principal issues of PHENIX2014~\cite{ppg100} was that an asymmetric system such as d$+$Au was required to resolve the difference between the Additive Quark Model---a color string model with the restriction of one color string per constituent-quark participant, or a maximum of 6 strings in d$+$Au compared to 3 in $p+p$---and the constituent-quark model which counts all the struck quarks in the Au target for producing particles. The result~\cite{ppg100} was that the AQM indicated a factor of 1.7 less \Et than the NQP, while the NQP is in excellent agreement with the data. The good news is that a calculation of the NQP model for d$+$Au with the Empirical Recentered method gives a result which is essentially indistinguishable from the PHENIX2014~\cite{ppg100} NQP result and the data (Fig.~\ref{fig:DVP}c). 
\subsection{Status of the two-component ansatz as an empirical proxy for \Nqp}
\label{app:proxy}
It has been popular since PHENIX~\cite{Adcox:2000sp}, inspired by the previous article in the same journal~\cite{WangGyulassyPRL86}, fit their measurement of the multiplicity distribution in Au$+$Au collisions at \sqsn=130 GeV to the equation:
\begin{equation}
d\Nch^{AA}/d\eta=d\Nch^{pp}/d \eta\ [(1-x)\,\mean{\Npart}/2 +x\,\mean{\Ncoll} ]
\label{eq:crazy}
\end{equation}
with $x\approx 0.16\pm 0.06$. PHOBOS~\cite{PHOBOSPRC70} also found that their measurement of $d\Nch^{AA}/d\eta$ in Au$+$Au at \sqsn=200 GeV was consistent with Eq~\ref{eq:crazy} with $x=0.09$ and also noted that their data were consistent with a constant value of $x\approx 0.13\pm 0.05$ from \sqs=19.6 to 200 GeV, more recently extended to $x\approx 0.10-0.12$ from \sqsn=19.6 GeV to $\sqsn=2.76$ TeV (Pb+Pb)~\cite{GSFSPRC90}. In PHENIX2014~\cite{ppg100} it was shown  that the ansatz \mbox{$[(1-x)\,\mean{\Npart}/2 +x\,\mean{\Ncoll} ]$} in Eq.~\ref{eq:crazy} works because the particular linear combination of \Npart and \Ncoll turns out to be an empirical proxy for \Nqp and not because the \Ncoll term implies hard-scattering in \Nch and \Et distributions---the ratio of $\mean{\Nqp}$/ansatz with $x=0.08$ varied by less than 1\% over the centrality range 0-80\% in 5\% bins. Given the new centered methods of calculating the \Nqp, does this relation still hold? 

\begin{figure}[ht] 
      \centering
      \includegraphics[width=0.94\linewidth]{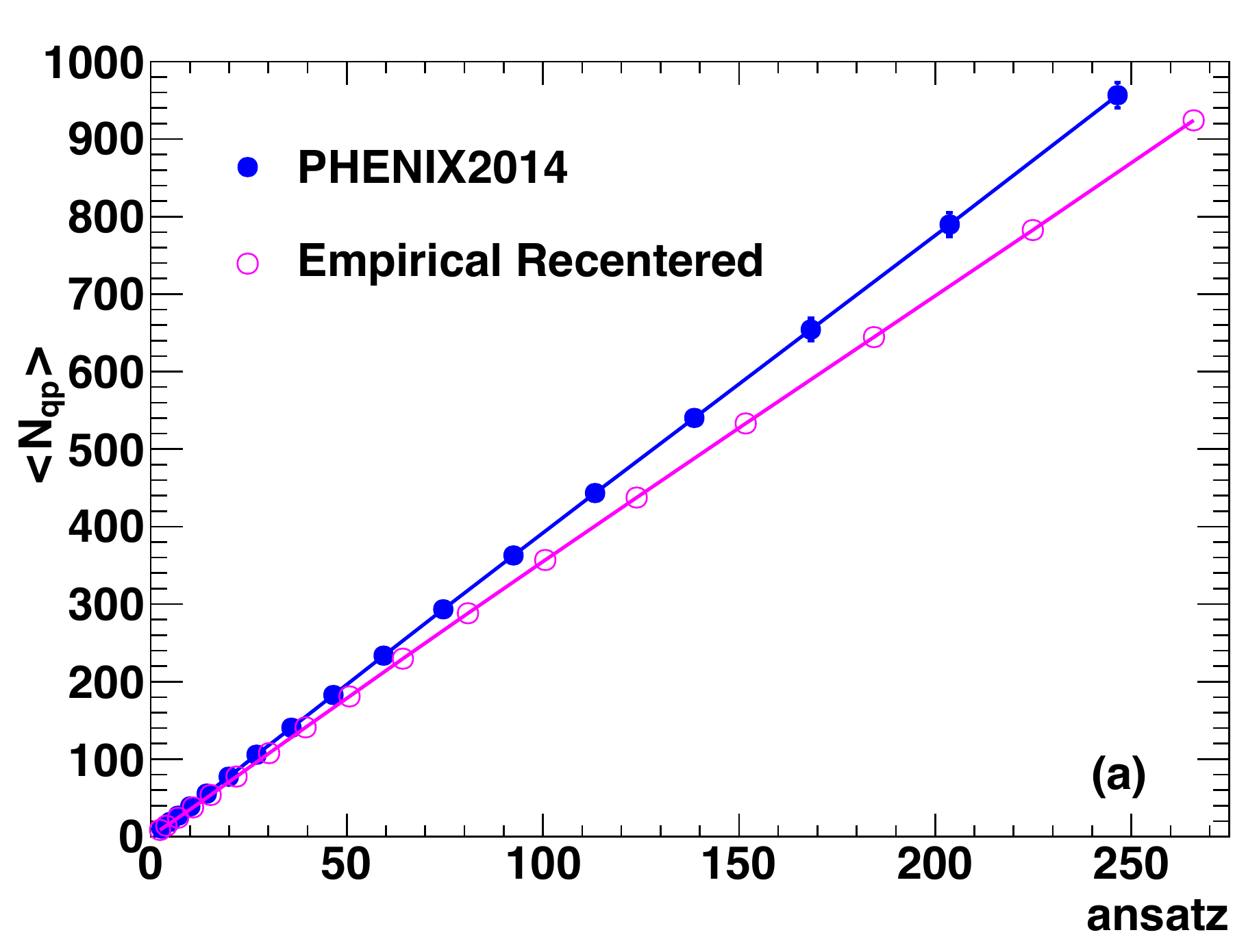}
      \includegraphics[width=0.98\linewidth]{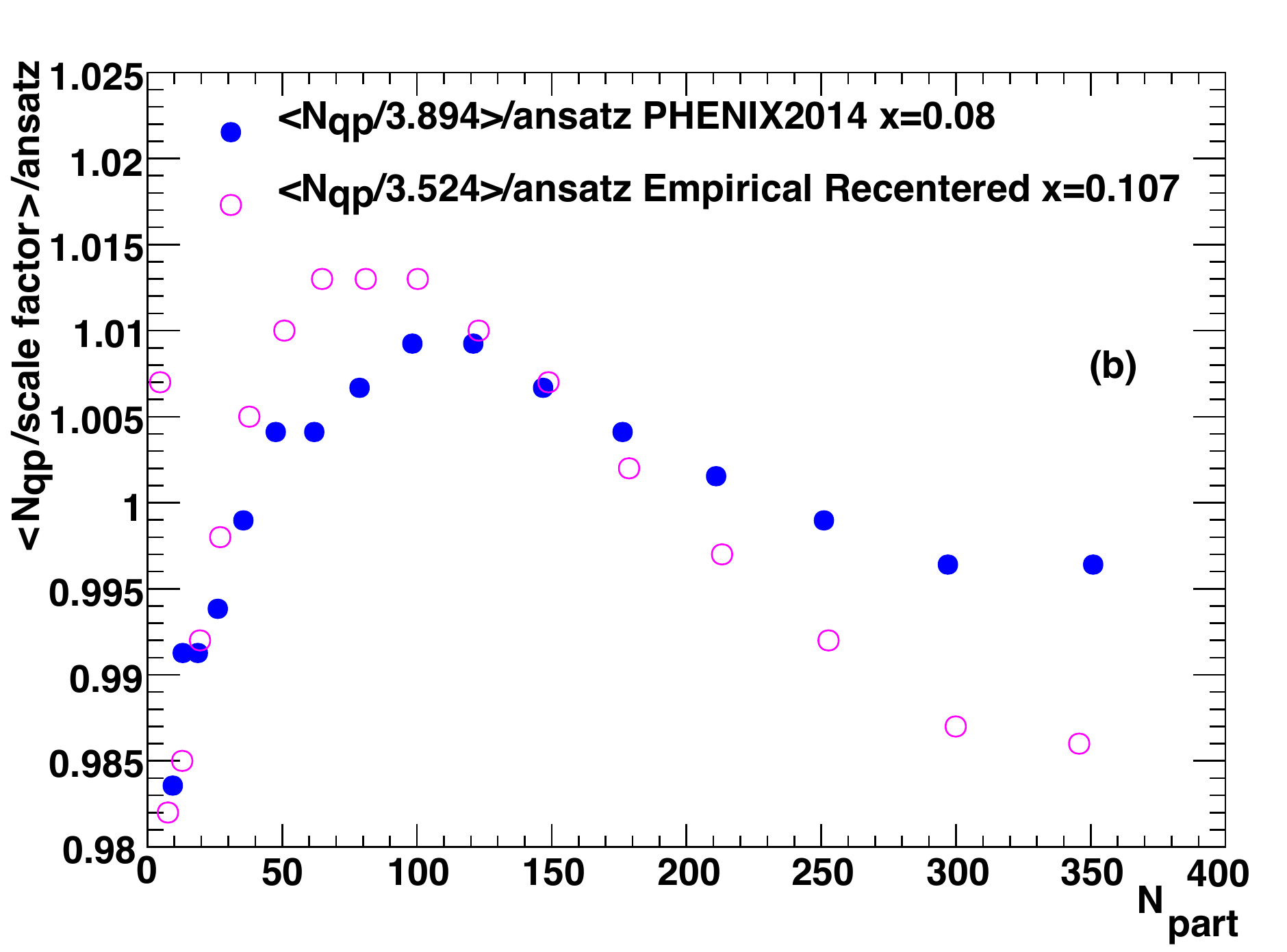}
            \caption[]{a) Plot of $\mean{\Nqp}$ vs. ansatz, $[(1-x)\,\mean{\Npart}/2 +x\,\mean{\Ncoll} ]$, with $x=0.107$, from Table~\ref{tab:auau200NcNpNqRatio}, compared to the PHENIX2014 result~\cite{ppg100}, with $x=0.08$  b) Plot of ratio of $(\mean{Nqp}/3.524)$/ansatz, where 3.524 is the average of $\mean{Nqp}$/ansatz for the Empirical Recentered method over the entire centrality range 0-92\% in 5\% bins, compared to the PHENIX2014 result~\cite{ppg100} with average of $\mean{Nqp}$/ansatz =3.894  }
      \label{fig:proxyproofD}
   \end{figure}

Figure~\ref{fig:proxyproofD}a shows the linearity of \Nqp vs. the ansatz in the Empirical Recentered method compared to the PHENIX2014~\cite{ppg100}  result, with the best values of $x$ in each case; and Fig.~\ref{fig:proxyproofD}b the  deviations of $\mean{\Nqp}$/ansatz/$\mean{\mean{\Nqp}/\rm ansatz}$ from 1.00 on an expanded scale (bottom) which are $\lsim 1.0$\% for Ref.~\cite{ppg100} and $\lsim 1.5$\% for the Empirical Recentered method.  This shows that the NQP model with either the PHENIX2014~\cite{ppg100} or the Empirical Recentered method are consistent from $p+p$ to Au$+$Au collisions and so actually work better than the ansatz (which does not extrapolate back to the $p+p$ value~\cite{ppg100}). 

\section{Conclusion}   
The three centered NQP calculations do not deviate from the PHENIX2014~\cite{ppg100} calculation of the \sqsn=200 GeV Au$+$Au \Et distribution by more than $1-1.5$ standard deviations of their systematic uncertainties. The surprising result is that the most sophisticated but computing intensive Explicit Joint method comes out virtually on top of the simplest Planar Polygon method for the Au$+$Au calculation rather than in agreement with the Empirical Recentered method. The Empirical Recentered method best solves the problem of maintaining the nucleon center and the correct charge distribution while keeping all the generated quark-triplets. 

The agreement of the new centered methods with the Au$+$Au data is within $1\sigma$ for the Empirical Recentered calculation (Fig.~\ref{fig:DVP}b) and $1.5\sigma$ for the Planar Polygon calculation (Fig.~\ref{fig:Poly}b); and the new calculations agree with PHENIX2014 calculation~\cite{ppg100} (Fig.~\ref{fig:allcalcs}) to within $1.2\sigma$ of its uncertainty for the Empirical Recentered method and $1.8\sigma$ for the Planar Polygon method. For d$+$Au, the NQP calculation with the Empirical Recentered method gives a result that is essentially indistinguishable from the PHENIX2014~\cite{ppg100} NQP result and the data (Fig.~\ref{fig:DVP}c).

The Empirical Recentered approach (Sec.~\ref{sec:DVPmodel}) was already used in the most recent PHENIX study~\cite{Adare:2015bua} of \Nqp scaling of midrapidity charged particle multiplicity, \dNch, and \dEt and did show a difference of $\lsim 2$\% between \dNchNormQ in PHENIX2014~\cite{ppg100} and the Empirical Recentered value~\cite{Adare:2015bua} as well as a similar small difference of $\mean{\dEt}/\Nqp = 0.617 \pm 
0.023$~GeV in PHENIX2014~\cite{ppg100} compared to the Empirical Recentered method with 
$\mean{\dEt}/\Nqp = 0.629 \pm 0.021$~GeV~\cite{Adare:2015bua}. Regarding the ansatz (Eq.~\ref{eq:crazy}), the NQP model with either the PHENIX2014~\cite{ppg100} or the Empirical Recentered method are consistent from $p+p$ to Au$+$Au collisions to within 1.5\% and in fact work better than the ansatz which does not extrapolate back to the $p+p$ value~\cite{ppg100}. 

 It is worth pointing out that the original PHENIX2014~\cite{ppg100} method and the Planar Polygon method agree with the best Empirical Recentered method to within 7\% lower \Et and 3\% higher \Et respectively for Au$+$Au. These two methods have an advantage in that it is straightforward to directly apply them to any number of quarks as well as any charge distribution so that they can quickly provide a bracket around the formally correct answer. The differences between the three new methods for the Au$+$Au calculations can be taken as a sort of ``modeling'' uncertainty---one that could improve with e.g. more information about constituent-quark correlations in a nucleon.  

\begin{acknowledgments}   
BNL research is supported by U.~S.~Department of Energy, Contract No. {DE-SC0012704}.
ORNL research is supported by U.~S.~Department of Energy, Contract No. DE-AC05-00OR22725.
\end{acknowledgments} 

\appendix
\section{Full details of the NQP calculations with the new methods. } 
\subsection{Original weights with no efficiency correction}
The calculations of the relative probabilities $w_n$ for $n=2,3,4,5,6$ constituent-quark participants and $\sigma_{q+q}$ in $p+p$ collisions with inelastic cross section 42 mb are given in Table~\ref{table:qqwts} for the PHENIX2014 and the three new methods.  
   \begin{table}[!bht]
\caption{Original weights $w_n$ ($p_0=0$, $\epsilon_{p+p}\equiv 1-p_0=1.0$) for $p+p$ at $\sqrt{s}=200$ GeV for the PHENIX2014~\cite{ppg100} and three different new models of the positions of 3 constituent-quarks which preserve the nucleon center. Note that in Ref.~\cite{ppg100} $\sigma=9.36$ mb was used as the $q+q$  scattering cross section order to obtain a N+N $\sigma^{\rm inel}=42.0$ mb. The $q+q$ scattering cross sections for the other models to obtain N+N $\sigma^{\rm inel}=42.3$ mb are also indicated in the table as well as the $\mean{\Nqp}$ per $p+p$ collision ($\Npart=2$).  
\label{table:qqwts}}
\begin{ruledtabular}
\begin {tabular}{ccccccc}
$\Nqp$  &PX2014~\cite{ppg100} & PlanarP  &  ExplicitJ & EmpiricalR\\
\hline
1 Ê& 0.00000   & 0.00000    & 0.00000    & 0.00000 \\ 
2 Ê& 4.6510E-01 &0.606528 Ê  &0.603101 Ê & 0.54301\\    
3 Ê& 2.3789E-01 &0.179538 Ê  &0.194968 Ê & 0.23186\\     
4 Ê& 1.6909E-01 &0.144047 Ê  &0.140203 Ê & 0.14464\\     
5 Ê& 9.4588E-02 &0.050957 Ê  &0.048279  Ê& 0.06351\\    
6 Ê& 3.3332E-02 &0.018930  Ê &0.013449 Ê & 0.01698\\   
\hline
sum  &1.0000002 & 1.000000    &1.000000  &1.000001\\
\hline
$\sigma_{q+q}$ mb & 9.36 & 7.72 & 8.15 & 8.17 \\
\hline
$\mean{\Nqp}$     &2.993 & 2.696&2.674 & 2.780\\
 \end{tabular}    \end{ruledtabular} 
\end{table}
One conclusion from Table~\ref{table:qqwts} is that $\mean{\Nqp/\Npart}$ only reduces from 1.50 in PHENIX2014 to 1.35 (Planar Polygon), 1.34 (Explicit Joint) and 1.39 (Empirical Recentered), i.e. a reduction from 1.5 to 1.4. 
For Au$+$Au, the original Au$+$Au weights, $w_n$, for the PHENIX2014~\cite{ppg100}, Planar Polygon, Explicit Joint and Empirical Recentered methods are presented in Table~\ref{table:AuAuwts}.   \begin{table}[hbt]
   	\caption{Original weights $w_n$ ($p_0=0$, $\epsilon_{p+p}\equiv 1-p_0=1.0$) for Au$+$Au at $\sqrt{s}=200$ GeV for the PHENIX2014 and three different new models of the positions of 3 constituent-quarks which preserve the nucleon center. \label{table:AuAuwts}}
   \begin{ruledtabular}
\begin {tabular}{ccccc}
$\Nqp$  &PX2014~\cite{ppg100} & PlanarP  &  ExplicitJ & EmpiricalR\\
\hline
1 Ê& 0.00000 & 0.00000 &0.00000 & 0.00000\\ 
2 Ê& 6.1275E-02 &5.0260E-02Ê  &5.4304E-02Ê&4.5933E-02\\    
3 Ê& 2.0448E-02 &1.9083E-02  &2.1482E-02   &2.1747E-02\\     
4 Ê& 2.0923E-02 &2.4041E-02Ê &2.6028E-02Ê &2.2329E-02\\     
5 Ê& 1.7606E-02 &1.7433E-02Ê& 1.8562E-02&1.8350E-02\\    
6 Ê& 1.5731E-02 &1.6620E-02Ê& 1.6977E-02Ê&1.6397E-02\\ 
7  & \ldots    & \ldots     & &\dots  \\
\hline
 \end{tabular}    \end{ruledtabular} 
\end{table}

\subsection{Weights corrected for $p+p$ efficiency, $\epsilon_{p+p}=1-p_0^{p+p}=0.647\pm 0.065$}

The method for the calculation of the \Et distribution from an $A$+$B$ 
reaction in a given detector is illustrated for the NQP or number of 
quark participants model. It has been discussed in detail in Ref.~\cite{ppg100} where it was emphasized that 
the key experimental issue is the linearity of the detector response to multiple collisions (better than 1\% in Ref.~\cite{ppg100}), and the stability of the response for the different $A$+$B$ combinations and run periods used in the analysis. The acceptance of the detector is taken into account by making a correction for the probability, $p_0$, of measuring zero \Et for an N+N inelastic collision, which is determined from the ratio of the measured $p+p$ cross section in the detector to the known inelastic cross section at \sqs=200 GeV~\cite{ppg100} and propagated to the $q+q$ collisions, with $\epsilon_{p+p}=1-p_0^{p+p}=0.647\pm 0.065$ for the PHENIX measurement~\cite{ppg100}.    

The \Et distribution is equal to the sum:
\begin{equation}
\bigg({d\sigma\over d\Et}\bigg)_{\rm NQP} = \sigma_{BA} \sum^{\rm N_{\rm max}}_{n=1} 
w_n\, P_n(\Et) 
\label{eq:wpnm}
\end{equation}
where $\sigma_{BA}$ is the measured $A$+$B$ cross section in the detector, 
$w_n$ is the relative probability for $n$ quark participants in the $A$+$B$ 
reaction with minimum value $n=2$ and maximum value $n=\rm N_{\rm max}$, and $P_n(\Et)$ is the 
calculated \Et distribution on the detector from $n$ independent quark participants. If $f_1(\Et)$ is the measured \Et spectrum on the detector for 
a quark participant that gives a nonzero \Et, and $p_0$ is the probability 
for a quark participant to produce no signal in the detector (zero \Et), 
then the correctly normalized \Et distribution for one quark participant is:
\begin{equation}
P_1(\Et)= (1-p_0)f_1(\Et) +p_0 \delta(\Et),
\label{eq:P1}
\end{equation}
where $\delta(\Et)$ is the Dirac delta function and $\int f_1(\Et)\, 
d\Et=1$.  $P_n(\Et)$ (including the $p_0$ effect) is obtained by 
convoluting $P_1(\Et)$ with itself $n-1$ times
\begin{equation}
P_n(\Et) = \sum ^n_{i=0} {{n!} \over {(n-i)!\ i!} } \, 
p_0^{n-i} (1-p_0)^i f_i(\Et) 
\label{eq:wpnm2}
\end{equation}
where $f_0(\Et)\equiv\delta(\Et)$   
and $f_i(\Et)$ is the $i$-th convolution of $f_1(\Et)$:
 \begin{equation}
f_i (x)=\int_0^x dy\, f_1(y)\,f_{i-1}(x-y)\;\;\; . \label{eq:defcon} 
\end{equation}
Substituting Eq.~\ref{eq:wpnm2} into Eq.~\ref{eq:wpnm} and 
reversing the indices gives a form that is less physically 
transparent, but considerably easier to compute: 
\begin{equation}
\bigg({d\sigma\over d\Et}\bigg)_{\rm NQP} = \sigma_{BA} \sum^{\rm N_{\rm max}}_{i=2} 
{w'}_i(p_0)\, f_i(\Et) 
\label{eq:wpnm3}
\end{equation}
where 
\begin{equation}
{w'}_i(p_0) = (1-p_0)^i\, \sum ^{\rm N_{\rm max}}_{n=i} {{n!} \over {(n-i)!\ i!} } \,
p_0^{n-i}\, w_n,  
\label{eq:wpnm4}
\end{equation}
which represents the weight (or relative probability) for $i$ convolutions 
of the measured $f_1(\Et)$ to contribute to the \Et spectrum in an $A$+$B$ 
collision, and where the terms with ${w'}_{i=0}(p_0)$ and ${w'}_{i=1}(p_0)$ in Eq.~\ref{eq:wpnm3} 
are left out because they represent the case when no signal is observed in 
the detector for an $A$+$B$ collision since there must be at least 2 quark participants i.e. 
\begin{equation}
{w'}_{i=0}(p_{0_{\rm NQP}})+{w'}_{i=1}(p_{0_{\rm NQP}})=p_0^{BA}.
\label{eq:p0conv}
\end{equation} 

For a $p+p$ collision this means that Eq.~\ref{eq:p0conv} for $p_0^{BA}\equiv p_0^{p+p}=1-\epsilon_{p+p}$=0.353  must be solved in order to find $p_{0_{\rm NQP}}$ and $\epsilon_{\rm NQP}=1-p_{0_{\rm NQP}}$. This is done in a spreadsheet and then the corrected weights ${w'}_{i}(p_{0_{\rm NQP}})$ for $i=2,3,4,5,6$ quark participants are calculated for $p+p$ collisions using these $p_{0_{\rm NQP}}$ in Eq.~\ref{eq:wpnm4}. 

The results for $\epsilon_{\rm NQP}$ for PHENIX2014~\cite{ppg100} and the three new methods along with the calculated corrected weights for $p+p$ collisions are presented in Table~\ref{table:corrwts}.  A test of the calculations is that the sum of the corrected weights ${w'}_{i}(p_{0_{\rm NQP}})$ for PHENIX2014 and the three new methods should equal the $p+p$ efficiency, the probability to get a non-zero $\Et$ on a $p+p$ collision, $\epsilon_{p+p}=1-p_0^{p+p}=0.647$, which works well.

   \begin{table}[htb]
   \caption{Corrected weights ${w'}_{i}$ for $p+p$, at $\sqrt{s}=200$ GeV. Note that $\epsilon_{p+p}=1-p_0^{p+p}$ is the sum of the corrected weights in the column and should equal the probability to get a non-zero $\Et$ on a $p+p$ collision, which should and does equal 0.647. 
\label{table:corrwts}}
\begin{ruledtabular}
  \begin{tabular}{cccccc}
\hline
$\Nqp$  &PX2014~\cite{ppg100} & PlanarP  &  ExplicitJ & EmpiricalR\\
\hline                                                                             
  &$\epsilon_{\rm QP}$=0.659&$\epsilon_{\rm QP}$=0.7074&$\epsilon_{\rm QP}$=0.7083&$\epsilon_{\rm QP}$=0.690\\
  \hline  
1    &0.0000E+00&0.00000& 0.0000E+00&0.0000E+00\\
2    &3.7795E-01&0.42684& 4.3082E-01&4.1102E-01\\
3    &1.7296E-01&0.14204& 1.4438E-01&1.5846E-01\\
4    &7.3126E-02&0.06083& 5.7330E-02&6.0648E-02\\
5    &2.0168E-02&0.01491& 1.2803E-02& 1.4873E-02\\
6    &2.7177E-03&0.00237& 1.6982E-03& 1.8326E-03\\
\hline
$\epsilon_{p+p}$ &6.46922E-01&0.64699& 6.47031E-01 & 6.46834E-01\\
\hline
\hline
 \end{tabular}    \end{ruledtabular} 
\end{table}

The corrected weights for Au$+$Au are given Table~\ref{table:AuAuCorrwts}.
   \begin{table}[htb]
   \caption{Corrected weights ${w'}_{i}$ for Au$+$Au, at $\sqrt{s}=200$ GeV. Note that $\epsilon_{\rm Au+Au}$ is the sum of the corrected weights in the column and is the probability to get a non-zero $\Et$ on an Au$+$Au collision. 
\label{table:AuAuCorrwts}}

\begin{ruledtabular}   \begin{tabular}{ccccc}
\hline
$\Nqp$  &PX2014~\cite{ppg100} & PlanarP  &  ExplicitJ & EmpiricalR\\
\hline                                                                             
  &$\epsilon_{\rm QP}$=0.659&$\epsilon_{\rm QP}$=0.7074&$\epsilon_{\rm QP}$=0.7083&$\epsilon_{\rm QP}$=0.690\\
  \hline  
1    &0.0000E+00&0.0000E+00& 0.0000E+00& 0.0000E+00\\
2    &4.7412E-02&4.3307E-02&4.7077E-02 & 4.1954E-02\\
3    &2.6998E-02&2.7143E-02&2.9262E-02 & 2.7895E-02\\
4    &2.3079E-02&2.3927E-02&2.5123E-02 & 2.3840E-02\\
5    &1.9473E-02&1.9816E-02&2.0473E-02 & 1.9966E-02\\
6    &1.6840E-02&1.7169E-02&1.7710E-02 & 1.7233E-02\\
7  & \ldots    & \ldots     & &\dots  \\
\hline
$\epsilon_{\rm Au+Au}$& 0.956& 0.968&  0.966      & 0.968\\
 \end{tabular}    \end{ruledtabular} 
\end{table}

\begin{figure}[ht] 
      \centering
      \includegraphics[width=0.94\linewidth]{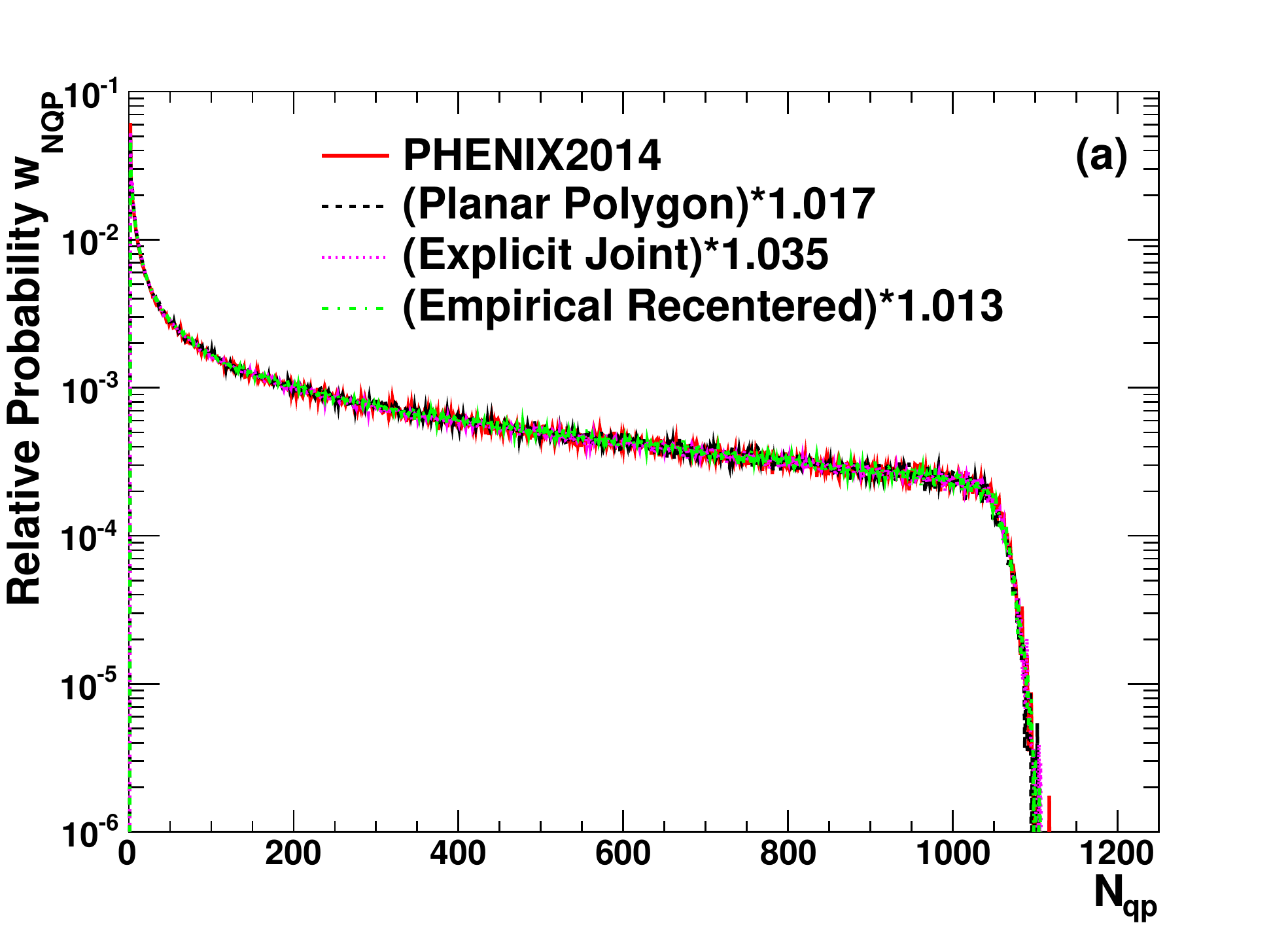}
      \includegraphics[width=0.94\linewidth]{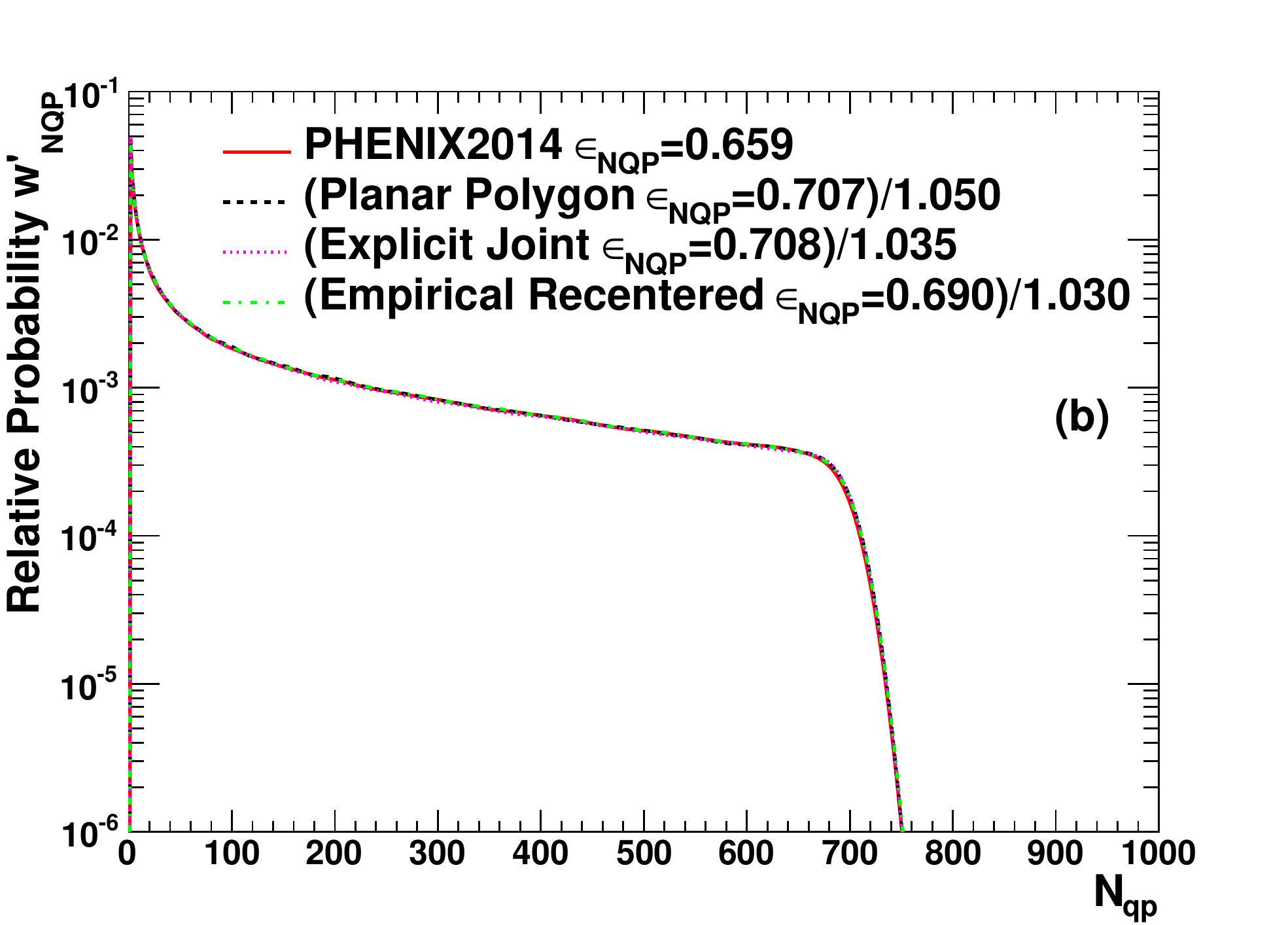}
            \caption[]{Distribution of the Number of Quark Participants in Au$+$Au at \sqsn=200\ GeV scaled in \Nqp by the indicated amounts so that they overlap the PHENIX2014 distribution~\cite{ppg100}: a) Orignial weights $w_{\rm NQP}$; b) Corrected weights $w'_{\rm NQP}$.  }
      \label{fig:AuAuwts}
   \end{figure}

The original (Table~\ref{table:AuAuwts}) and corrected (Table~\ref{table:AuAuCorrwts}) weights for Au$+$Au are plotted in Fig.~\ref{fig:AuAuwts}. The weights are scaled in \Nqp by the amounts indicated for the three new methods and all lie on top of the PHENIX2014~\cite{ppg100} distributions when scaled up in \Nqp by 1.3\% to 3.5\% for the original weights (Fig.~\ref{fig:AuAuwts}a) and scaled down in \Nqp by 3\% to 5\% for the corrected weights. 
(Fig.~\ref{fig:AuAuwts}b). 


\subsection{Calculation of the \Et distribution of a quark participant for the PHENIX2014 data~\cite{ppg100}}
\label{sec:norm}
The \Et distributions for $p+p$, and Au$+$Au at $\sqsn=200$ GeV~\cite{ppg100} were shown in Fig~\ref{fig:PXppg100}. They have been corrected to hadronic \Et in $\delta\phi=2\pi$ and $\delta\eta=1.0$.
The total number of counts in the distributions (including the counts with zero \Et which are in the lowest bin) sum up to the number of Beam Beam trigger (BBC) counts. The distributions are then normalized so that the integral is unity and represents the yield per BBC count per GeV, $dY/d\Et$. The lowest bin also contains counts with non-zero \Et. 
 The \Et distribution for a $p+p$ collision is first fit to a single Gamma distribution which integrates to $Y_{\Gamma}^{pp}=0.93349$ (only including the non-zero counts in the lowest bin) which is the observed yield per BBC count (see Table ~\ref{table:Gammafits}). Then the BBC counts with zero in the lowest bin are removed so that only the fitted non-zero counts remain, to give the observed \Et distribution in $p+p$ which is used for all the following fits.

The \Et distribution of a quark participant is derived by assuming that the measured $p+p$ \Et spectrum in the experiment is composed of the sum of the \Et distributions emitted independently by $n=2,3,4,5,6$ constituent-quark participants weighted by their corrected probabilities $w'_n$ given in Table~\ref{table:corrwts}.  
This is done by fitting Eq.~\ref{eq:wpnm3} to the observed $p+p$ data, where $f_1(\Et)$ is taken as a Gamma distribution:\begin{equation}
f_1(x)=\frac{b}{\Gamma(p)} (bx)^{p-1} e^{-bx},
\label{eq:gammadist}
\end{equation}
where \[ p>0,\quad b>0,\quad 0\leq x\leq \infty \quad ,\] $\Gamma(p)$ is
is the Gamma function, which equals $(p-1)!$ if $p$ is an integer, and 
$\int_0^\infty f_1(x)\, dx=1$. 

There are two reasons for this: i) in general the shape of \Et distributions 
in $p$$+$$p$ collisions is well represented by the Gamma distribution; and ii)
the $n$-fold convolution (Eq.~\ref{eq:defcon}) is analytical i.e. $p\rightarrow np$ while $b$ remains unchanged. 
\begin{equation}
f_n(x)=\frac{b}{\Gamma(np)} (bx)^{np-1} e^{-bx},
\label{eq:gammaconv}
\end{equation}

For this fit, which we call the deconvolution of the $p+p$ \Et distribution to its constituent-quark components (represented by Gamma distribution parameters $b$ and $p$),  
the $\sigma_{BA}$ in Eq.~\ref{eq:wpnm3} is replaced by $Y_{\Gamma}^{\rm fit}$. Also the integral of the sum of the weights times the normalized Gamma distributions which multiplies $Y_{\Gamma}^{\rm fit}$ is not unity but is equal to $\epsilon_{p+p}=0.647$, so that $Y_{\Gamma}^{\rm fit}\times \epsilon_{p+p}$ should equal the simple integral of the $p+p$ distribution, $Y_{\Gamma}^{pp}$, obtained by the fit of a single Gamma distribution to the measured $p+p$ \Et distribution (Table~\ref{table:Gammafits}, NCOLL model=1 $p+p$ collision). 

The derived parameters $b$ and $p$ of the \Et distribution of a constituent-quark for all four models are given in Table~\ref{table:Gammafits} and the $Y_{\Gamma}^{\rm fit}\times \epsilon_{p+p}$ agrees with the simple integral of the $p+p$ measured distribution to within 1.6\% with the new models in better agreement than the original PHENIX2014  calculation.

   \begin{table*}[!t]
   \caption{Parameters $b$, $p$ of the element indicated from the fit to $p+p$ data, cut for $\Et<13.3$ GeV. $Y_{\Gamma}^{\rm fit}$ is the fitted integral of the $p+p$ distribution. For NCOLL the fit is to a single $\Gamma$ distribution from which $\epsilon_{p+p}$ is calculated~\cite{ppg100}. For the NQP models the fits are the deconvolution of elements with weights ${w'}_{i}$ which do not sum to unity but sum to $\epsilon_{p+p}=0.647$ so that $Y_{\Gamma}^{\rm fit}\times \epsilon_{p+p}$ should equal $Y_{\Gamma}^{pp}=0.93349$---the actual values are 0.948 (PHENIX2014) and 0.945  (Planar Polygon), 0.945 (Explicit Joint), 0.946 (Empirical Recentered)  a good check (within 1.6\%,1.3\%, 1.3\% and 1.3\% respectively). The $ \chi^2$/dof of the deconvolution fits to the $p+p$ \Et distribution are 4907/17 (PHENIX2014) and 4841/17 (Planar Polygon), 4818/17 (Explicit Joint), 4836/17 (Empirical Recentered) which for the three centered methods are better than the single $\Gamma$ fit 4866/17 (NCOLL). \label{table:Gammafits}}
\begin{ruledtabular}   \begin{tabular}{cccccccc}
\hline
Model & $\epsilon_{\rm element}$ & element &$Y_{\Gamma}^{\rm fit}$ & $b$ (GeV)$^{-1}$ & $p$ & $\langle E_{T}\rangle^{\rm elem}_{\rm fit}$ (GeV)\\
\hline
\hline
NCOLL & 0.647  & p+p   & 0.93349   & 1.8259/6.68 & 0.72359 & 0.396 $\times 6.68$ \\
PHENIX2014   & 0.659  & 1 QP  & 1.4659  & 1.9986/6.68 & 0.29740 & 0.1488 $\times 6.68$ \\
Planar Polygon & 0.707  & 1 QP  & 1.4613  & 1.9935/6.68 & 0.31114 & 0.1561 $\times 6.68$\\
Explicit Joint      & 0.708  & 1 QP  & 1.4605  & 1.9836/6.68 & 0.31251 & 0.1575 $\times 6.68$\\
Empirical Recentered  & 0.690  & 1 QP  & 1.4626  & 1.9867/6.68 & 0.30726 & 0.1547 $\times 6.68$\\
\hline
 \end{tabular}    \end{ruledtabular} 
\end{table*}
\normalsize

\begin{figure}[!h] 
      \centering 
      \includegraphics[width=0.94\linewidth]{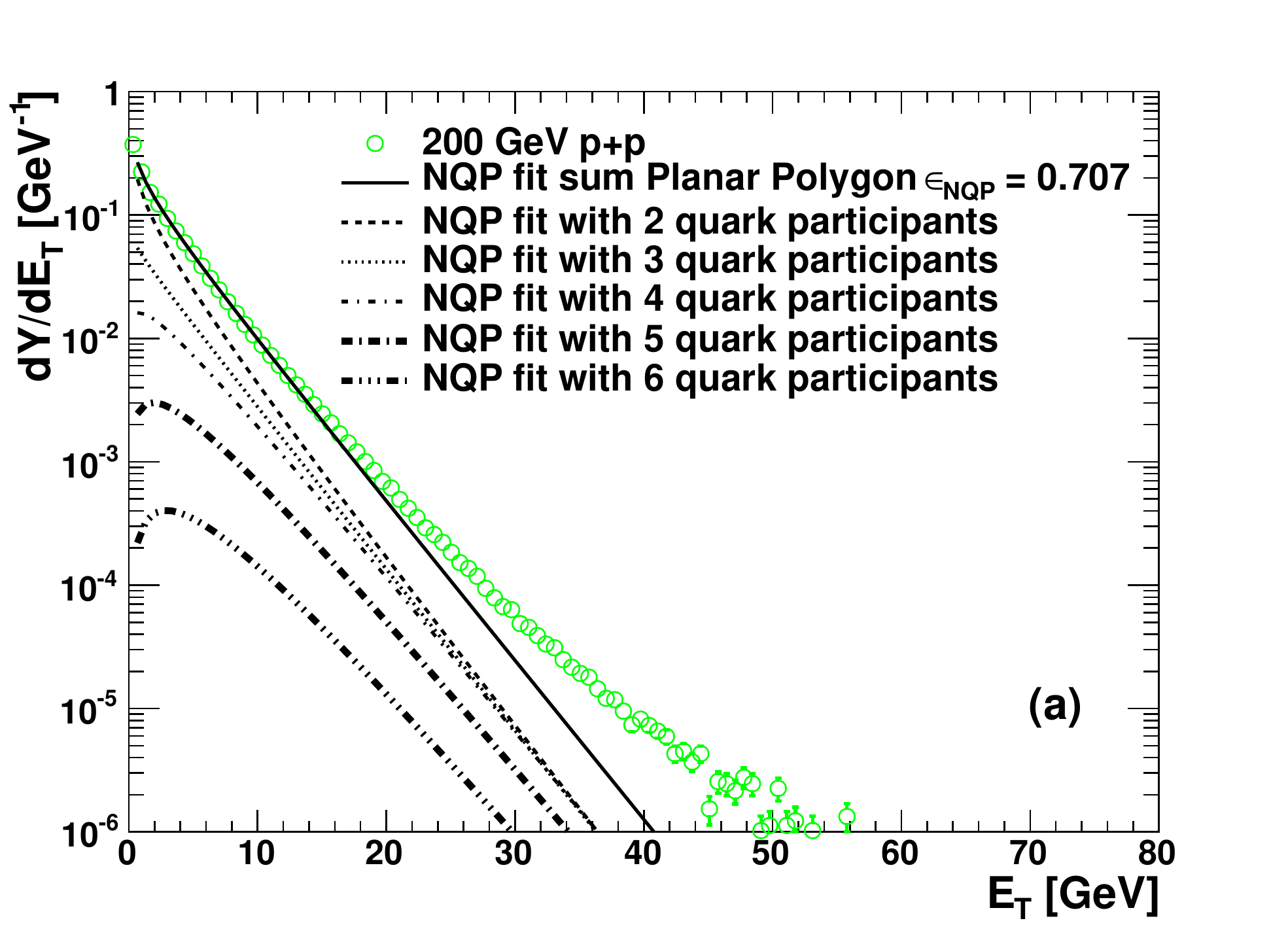}
      \includegraphics[width=0.94\linewidth]{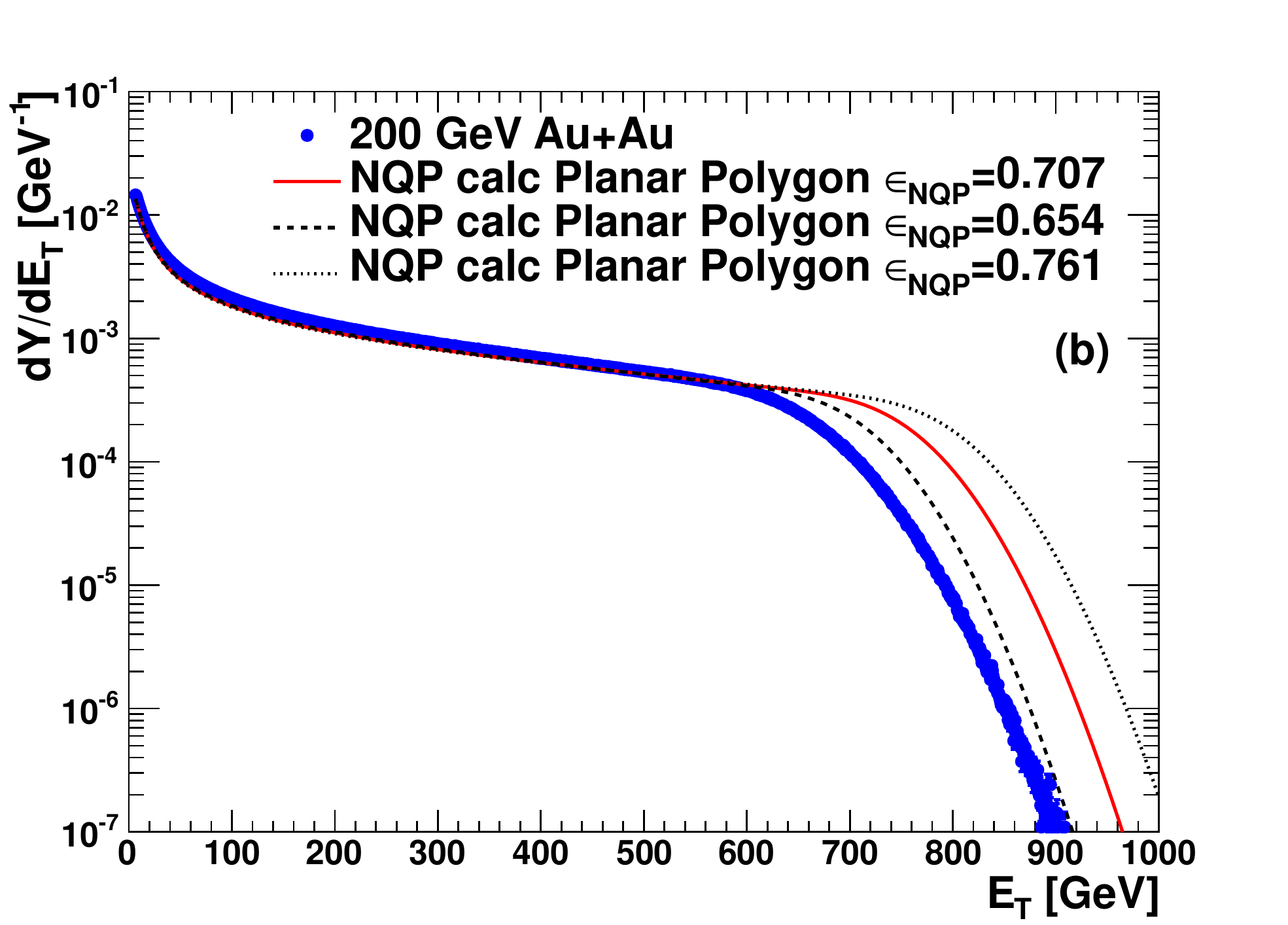} 
      \caption{\Et distributions at $\sqrt{s_{\rm NN}}=200$ GeV calculated in the Number of Quark Participants (NQP) model  from the Planar Polygon method, with  $\epsilon_{\rm NQP}=1-p_{0_{\rm NQP}}=0.707$. a) NQP calculation of the $p+p$ data with the parameters $p$ and $b$ for the \Et distribution of 1 QP from the deconvolution fit to the same data (Table~\ref{table:Gammafits}) where the thin lines shown are the \Et distributions for 2,3,4,5 and 6 QP weighted by $w'_n$ from Table~\ref{table:corrwts} and the thick line is the sum. b) Au+Au calculation with systematic uncertainties.}
      \label{fig:Poly}
   \end{figure}

\begin{figure}[!h] 
      \centering
      \includegraphics[width=0.94\linewidth]{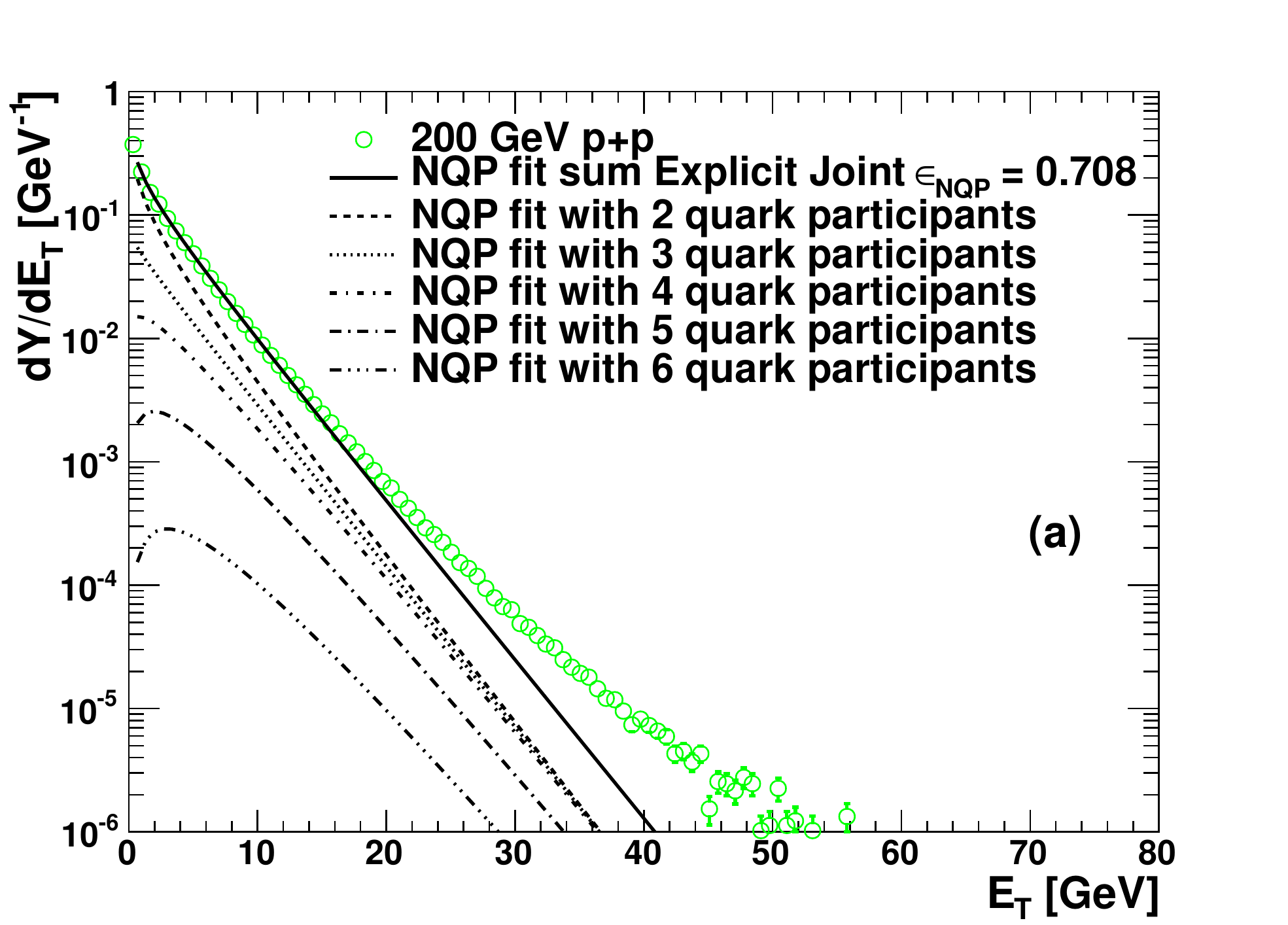}
      \includegraphics[width=0.94\linewidth]{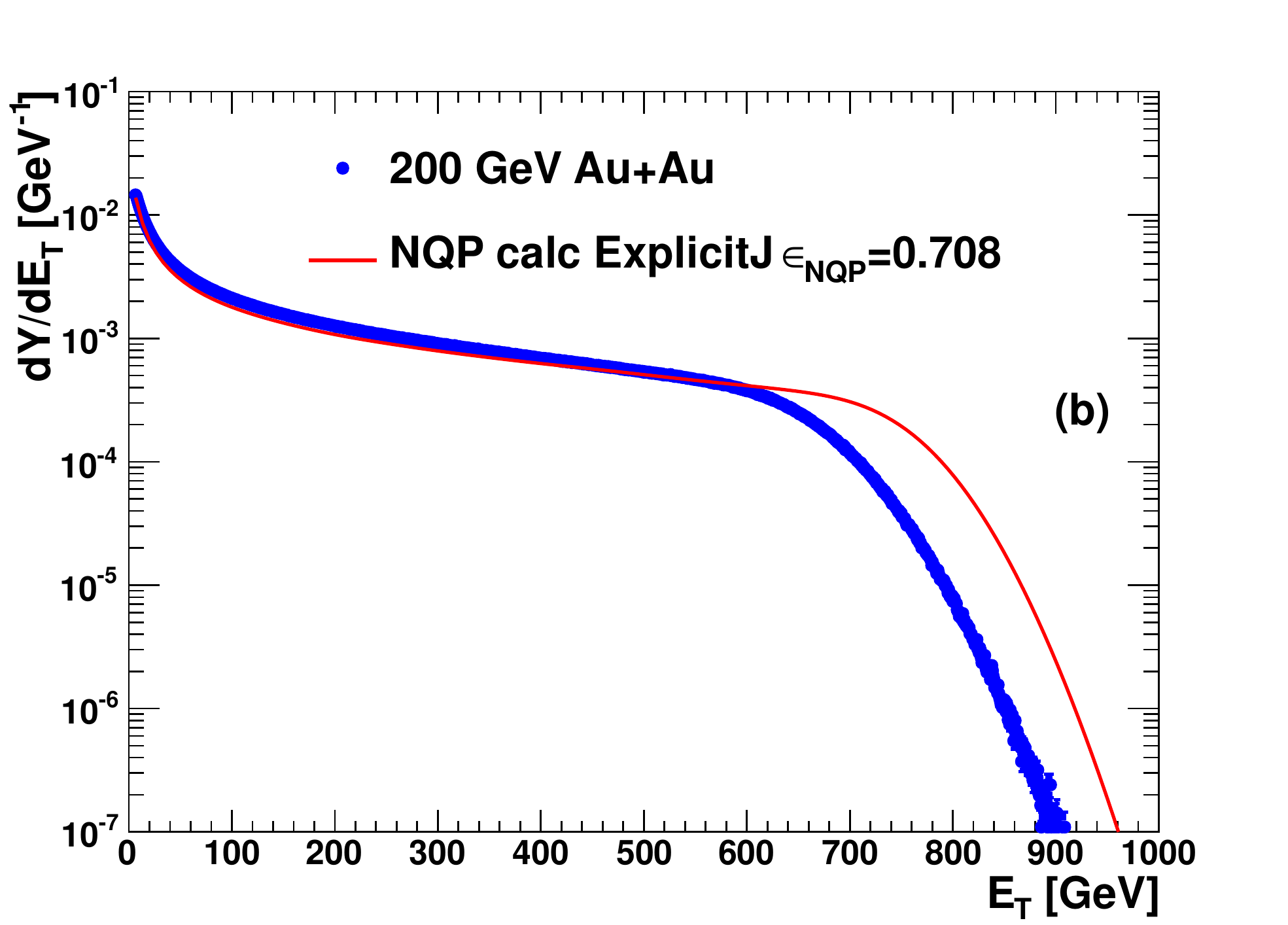} 
      \caption{\Et distributions at $\sqrt{s_{\rm NN}}=200$ GeV calculated in the Number of Quark Participants (NQP) model  from the Explicit Joint method, with  $\epsilon_{\rm NQP}=1-p_{0_{\rm NQP}}=0.708$. a) NQP calculation of the $p+p$ data with the parameters $p$ and $b$ for the \Et distribution of 1 QP from the deconvolution fit to the same data (Table~\ref{table:Gammafits}) where the thin lines shown are the \Et distributions for 2,3,4,5 and 6 QP weighted by $w'_n$ from Table~\ref{table:corrwts} and the thick line is the sum. b) Au+Au calculation.}
      \label{fig:m3}
   \end{figure}

The p$+$p fits and Au$+$Au calculations for the Planar Polygon (Sec.~\ref{sec:MJT}) and Explicit Joint (Sec.~\ref{sec:PWS}) methods are shown in Figs.~\ref{fig:Poly} and ~\ref{fig:m3} respectively.

   \begin{table}[!h]
\caption{Uncorrected and Corrected weights, $w_{i}$ and ${w'}_{i}$ for d$+$Au, at $\sqrt{s}=200$ GeV. 
\label{table:dAuWts}}

\begin{ruledtabular}   \begin{tabular}{ccccc}
\hline
$\Nqp$  & PX2014~\cite{ppg100}& PX2014~\cite{ppg100}  &EmpiricalR& EmpiricalR\\ 
\hline                                                                             
& uncorrected &$\epsilon_{\rm NQP}$=0.659& uncorrected &$\epsilon_{\rm NQP}$=0.690\\
  \hline  
1    &0.0000E+00&0.0000E+00& 0.0000E+00& 0.0000E+00\\
2    &8.6717E-02&9.1794E-02& 1.0191E-01& 1.0044E-01\\
3    &5.1578E-02&7.2569E-02& 5.4985E-02 & 7.3771E-02\\
4    &5.2929E-02&6.6353E-02& 5.7050E-02& 6.6407E-02\\
5    &4.7335E-02&6.0144E-02& 4.9220E-02& 5.9248E-02\\
6    &4.5134E-02&5.5165E-02& 4.5505E-02& 5.4320E-02\\
7  & \ldots    & \ldots     &\ldots &\ldots  \\
\hline
 \end{tabular}    \end{ruledtabular} 
\end{table}

\pagebreak
   
\subsection{d+Au and AQM vs NQP}
The uncorrected and corrected d$+$Au weights for PHENIX2014 and the Empirical recentered approach are given in Table~\ref{table:dAuWts}. The d$+$Au NQP calculation with the Empirical Recentered method previously shown in Fig.~\ref{fig:DVP}c gives a result that is essentially indistinguishable from the PHENIX2014~\cite{ppg100} NQP result and the data.


\subsection{Calculation of the ansatz}
In Table~\ref{tab:auau200NcNpNqRatio}, the centrality dependence of $\mean{N_{part}}$, $\mean{N_{qp}}$,  $\mean{\Ncoll}$ is presented, together with the ansatz calculated for $x=0.107$ in the Empirical recentered approach, which gives the best result. 
The average of the $\mean{\Nqp}/{\rm ansatz}=3.524$ over the entire centrality range in Au$=$Au from 0-92\% in 5\% bins and varies within $\pm 1.5$\% over this range, still an excellent result. Figure~\ref{fig:proxyproofD} showed the linearity of \Nqp vs. the ansatz compared to the PHENIX2014~\cite{ppg100}  result, with the best values of $x$ in each case  (top); and the $\lsim 1.5$\%  deviations of $\mean{\Nqp}$/ansatz/$\mean{\mean{\Nqp}/\rm ansatz}$ from 1.00 on an expanded scale (bottom).  

  The fact that the $\mean{\Nqp}$/ansatz ratio drops from an average of 3.524 for Au$+$Au collisions to 2.78 for $p$$+$$p$ collisions is also interesting. This is consistent with the PHOBOS~\cite{PHOBOSPRC70} result that a fit of Eq.~\ref{eq:crazy} to
$\mean{d\Nch^{AA}/d\eta}$ with $x=0.09$ leaving $\mean{d\Nch^{pp}/d\eta}$ as a free parameter gives the result $\mean{d\Nch^{pp}/d\eta}=2.70$ which is above the measured inelastic value of 2.29. The lower value of $\mean{\Nqp}$/ansatz for $p$$+$$p$ would then give a value of $2.70\times2.78/3.524=2.13$ for $\mean{d\Nch^{pp}/d\eta}$, much closer to the measured  value. This shows that the \Nqp model is consistent from from $p+p$ to Au$+$Au collisions and so actually works better than the ansatz. 

\begin{table*}[!thb]
\caption{\label{tab:auau200NcNpNqRatio}
Test of whether the ansatz, \mbox{$[(1-x)\,\mean{\Npart}/2 +x\,\mean{\Ncoll} ]$}, from Eq.~\ref{eq:crazy}, with $x=0.107$, is a proxy for $\mean{\Nqp}$. The errors are not quoted on $\mean{N_{part}}$, $\mean{N_{qp}}$,  $\mean{\Ncoll}$ because they are correlated Type C and largely cancel for the $\mean{\Nqp}$/ansatz ratio.   }
\begin{ruledtabular}   
\begin{tabular}{ccccccc}
\hline\hline
Centrality & $\mean{N_{part}}$ & $\mean{N_{qp}}$ & $\mean{\Ncoll}$ & ansatz & $\mean{\Nqp}$/ansatz& $\mean{\Nqp}$/ansatz/$\mean{\mean{\Nqp}/{\rm ansatz}}$ \\\hline
0-5\%&    345.7& 924.2& 1042.5& 265.9& 3.476& 0.986\\
5-10\% &  299.9& 782.6&  850.3& 224.9& 3.480& 0.987\\
10-15\% & 252.7& 644.5&  668.7& 184.4& 3.496& 0.992\\
15-20\% & 213.2& 533.1&  527.7& 151.7& 3.515& 0.997\\
20-25\% & 178.7& 437.5&  412.1& 123.9& 3.531& 1.002\\
25-30\% & 148.8& 356.9&  318.8& 100.6& 3.549& 1.007\\
30-35\% & 122.9& 288.2&  243.4&  80.9& 3.561& 1.010\\
35-40\% & 100.3& 229.6&  182.7&  64.3& 3.570& 1.013\\
40-45\% &  81.0& 180.9&  135.4&  50.7& 3.572& 1.013\\
45-50\% &  64.8& 141.0&   98.9&  39.5& 3.570& 1.013\\
50-55\% &  50.8& 107.7&   70.6&  30.2& 3.561& 1.010\\
55-60\% &  37.8&  77.5&   47.0&  21.9& 3.542& 1.005\\
60-65\% &  27.0&  53.7&   29.9&  15.3& 3.517& 0.998\\
65-70\% &  19.5&  37.7&   19.5&  10.8& 3.497& 0.992\\
70-75\% &  12.9&  24.3&   11.6&   7.0& 3.473& 0.985\\
75-80\% &   7.6&  14.0&    6.1&   4.0& 3.460& 0.982\\
80-92\% &   4.7&   8.6&    3.3&   2.4& 3.549& 1.007\\
\hline
$p+p$   &   2 &   2.78 &   1 &  1 & 2.78& 0.79\\       
\hline
 \end{tabular}    \end{ruledtabular} 
\end{table*}

\bibliography{CQ}
\end{document}